\documentclass[fleqn,usenatbib]{mnras}

\usepackage{afterpage}

\usepackage[T1]{fontenc}

\usepackage{graphicx}	
\usepackage{amsmath}	
\usepackage{amssymb}	
\usepackage{rotating}
\usepackage{pdflscape}
\usepackage[dvipsnames]{xcolor}
\usepackage{array, makecell}

\DeclareRobustCommand*{\ha}{H$\alpha$}
\DeclareRobustCommand*{\hb}{H$\beta$}
\DeclareRobustCommand*{\hg}{H$\gamma$}
\DeclareRobustCommand*{\lya}{Ly$\alpha$}

\DeclareRobustCommand*{\msunyr}{M$_\odot$/yr}
\DeclareRobustCommand*{\hst}{{\it HST}}
\DeclareRobustCommand*{\ebvcon}{$E(B-V)_\text{cont}$}

\usepackage{newtxtext,newtxmath}

\title[Scrutiny of a young, metal-poor LAE at $z\approx 3.7$]
{Scrutiny of a very young, metal-poor star-forming \lya-emitter at $z\approx 3.7$}

\author[E. Iani et al.]{E. Iani$^{1}$\thanks{E-mail: iani@astro.rug.nl}, 
A. Zanella$^{2}$,
J. Vernet$^{3}$,
J. Richard$^{4}$,
M. Gronke$^{5}$,
F. Arrigoni-Battaia$^{5}$,
\newauthor
A. Bolamperti$^{6}$,
K. Caputi$^{1}$,
A. Humphrey$^{7}$,
G. Rodighiero$^{6}$, 
P. Rinaldi$^{1}$,
E. Vanzella$^{8}$\\
\\
$^{1}$Kapteyn Astronomical Institute, University of Groningen, 9700AV Groningen, The Netherlands\\
$^{2}$Istituto Nazionale di Astrofisica (INAF), Vicolo dell’Osservatorio 5, I-35122 Padova, Italy\\
$^{3}$European Southern Observatory, Karl Schwarzschild Straße 2, D-85748 Garching, Germany\\
$^{4}$Univ. Lyon, Univ Lyon1, ENS de Lyon, CNRS, Centre de Recherche Astrophysique de Lyon UMR5574, F-69230, Saint- Genis-Laval, France\\
$^{5}$Max-Planck-Institut fur Astrophysik, Karl-Schwarzschild-Str 1, D-85748 Garching bei München, Germany\\
$^{6}$Dipartimento di Fisica ed Astronomia, Università degli Studi di Padova, Vicolo dell’Osservatorio 3, I-35122 Padova, Italy\\
$^{7}$Instituto de Astrofísica e Ciências do Espaço - Centro de Astrofísica da Universidade do Porto, Rua das Estrelas, 4150-762, Porto, Portugal\\
$^{8}$Istituto Nazionale di Astrofisica (INAF), Osservatorio di Astrofisica e Scienza dello Spazio, via Gobetti 93/3, 40129 Bologna, Italy\\
}

\date{Accepted XXX. Received YYY; in original form ZZZ}

\pubyear{2022}

\begin{document}
\label{firstpage}
\pagerange{\pageref{firstpage}--\pageref{lastpage}}
\maketitle

\begin{abstract}
The origin of the Lyman-$\alpha$ (\lya) emission in galaxies is a long-standing issue: despite several processes known to originate this line (e.g. AGN, star formation, cold accretion, shock heating), it is difficult to discriminate among these phenomena based on observations. 
Recent studies have suggested that the comparison of the ultraviolet (UV) and optical properties of these sources could solve the riddle.
For this reason, we investigate the rest-frame UV and optical properties of A2895b, a strongly lensed \lya-emitter at redshift $z\sim 3.7$. 
From this study, we find that our target is a compact ($r_{n}\sim 1.2$~pkpc) star-forming (star formation rate $\simeq 11~{\rm M}_\odot$/yr) galaxy having a young 
stellar population. 
Interestingly, we measure a high ratio of the \hb\ and the UV continuum monochromatic luminosities (${\rm L(H\beta)/L(UV)}\simeq 100$).
Based on tracks of theoretical stellar models (\textsc{Starburst99}, \textsc{bpass}), we can only partially explain this result by assuming a recent ($\lesssim 10$ Myr), bursty episode of star-formation and considering models characterised by binary stars, a top-heavy initial-mass function (IMF) and sub-solar metallicities (${\rm Z} \lesssim 0.01~{\rm Z}_\odot$).
These assumptions also explain the observed low (C/O) abundance of our target ($\simeq 0.23 {\rm (C/O)}_\odot$).
By comparing the UV and optical datasets, we find that the \lya\ and UV continuum are more extended ($\times 2$) than the Balmer lines, and that the peak of the \lya\ is offset ($\simeq 0.6$~pkpc).
The multi-wavelength results of our analysis suggest that the observed \lya\ emission originates from a recent star-formation burst, likely taking place in an off-centre clump.

\end{abstract}




\section{Introduction}
\label{sec:intro}
In the hydrogen atom, whenever an electron falls from the first excitation level $2p$ to the ground state $1s$, a photon with an energy of 10.2~eV and a wavelength of 1215.67 \AA\ is emitted.
This ultraviolet (UV) transition is the brightest hydrogen emission and it is commonly referred to as Lyman-$\alpha$ (\lya) line \cite[][]{Lyman+06}.
Thanks to its brightness and the fact that hydrogen constitutes about 74 per cent of the baryonic matter in the Universe \cite[e.g.][]{Croswell+96,Carroll+06}, the \lya\ acts as a beacon for the detection of galaxies at intermediate/high-redshifts ($z\gtrsim 2-3$).
In fact, its UV rest-frame wavelength is shifted into the optical and near-infrared (NIR) at cosmological distances (cosmological redshift).

Galaxies detected through their \lya\ emission (and having a rest-frame \lya\ equivalent width EW$_0 \gtrsim 20$~\AA) are generally referred to as \lya-{\it emitters} \cite[LAEs, e.g.][]{Ouchi+20}.
In the last decades, studies have revealed that these systems have often compact morphologies with effective radii $r_e \sim 1$ pkpc \cite[e.g.][]{Pascarelle+96, Matthee+21, Pucha+22} and, typically, a disk-like radial surface-brightness (SB) profile with a S\'ersic index of $n_s \sim 1$ \cite[e.g.][]{Taniguchi+09, Gronwall+11}.
Whenever in the absence of an active galactic nucleus (AGN), LAEs are found to be low-mass (stellar mass $M_\star\sim 10^{7-9} M_\odot$), young (stellar ages $\sim 10$ Myr) star-forming galaxies (SFGs) with star formation rates ${\rm SFR} \sim 1-10~ M_\odot/{\rm yr}$ \cite[e.g.][]{Nakajima+12, Hagen+14, Hagen+16, Matthee+21, Pucha+22}.
With a specific star formation rate sSFR $\gtrsim 10^{-7} {\rm yr}^{-1}$ (${\rm sSFR}={\rm SFR}/M_\star$), \lya-emitters are generally starbursting systems.
As for the properties of their interstellar medium (ISM), LAEs are dust poor galaxies with stellar and nebular colour extinction values $E(B-V) \sim 0-0.2$ \cite[][]{Ono+10, Kojima+17}, and a gas-phase metallicity (derived from both strong lines and direct electron temperature T$_e$ methods) ${\rm Z}\sim 0.1 - 0.5~{\rm Z}_\odot$ \cite[e.g.][]{Finkelstein+11, Nakajima+12, Trainor+16, Kojima+17}.
Despite all these findings, the origin of their \lya\ emission is still debated.

Up to date, five major phenomena are generally invoked to explain \lya\ emission in and surrounding galaxies \cite[e.g.][]{Ouchi+20}: in-situ and/or ex-situ (i.e. from unresolved faint satellite galaxies, e.g. \citealt{Mas-Ribas+17}) star formation, AGN, shock heating due to outflows, cold accretion via gravitational cooling, and fluorescence, i.e. when ionising photons can escape their production area and reach and ionise pockets of ISM far from star-forming regions. 
Despite the fact that each one of these processes can originate the \lya\ line, the emission observed in galaxies is possibly driven by a combination of phenomena, each one dominant on a different scale \cite[see][and references therein]{Claeyssens+22}.
To discriminate between these different phenomena is, therefore, a demanding task.

Notwithstanding its brightness, the \lya\ line is complex to study since it strongly suffers from both dust extinction and resonant scattering.
While the main effect of dust is to erode the UV flux and to re-emit it at the infrared (IR) wavelengths, resonant scattering is a diffusive process where \lya\ photons do not escape freely from their production site but undergo a number of absorption and remission events due to intervening atoms of neutral hydrogen along the photons propagation line. 
The amount of scatterings strongly depends on the properties of the medium the \lya\ photons diffuse into, i.e. its neutral hydrogen column density, geometry and kinematics \cite[see, e.g.][and references therein]{Dijkstra+14}.
Besides, each scattering slightly alters the frequency of the \lya\ photons, as well as their direction of propagation \cite[][]{Osterbrock+62}.
Hence, the spectral characteristics of the emerging radiation (i.e. the spectral shape of the observed \lya) do not only encode the properties of the phenomena driving the emission of the line but also those of the scattering medium along the paths that offered least resistance to the diffusing photons \cite[e.g.][]{Dijkstra+16,Gronke+16}.
This makes observationally challenging determining what are the actual processes driving the \lya\ emission.

A potential way to tackle this problem is to investigate in depth the properties of the host galaxy and its surrounding medium, and possibly, spatially resolve them.
In particular, the  joint analysis of the SB profiles and spatial extent of optical hydrogen transitions (i.e. the Balmer lines), \lya\ and UV stellar continuum are believed to be a useful tool to disentangle the different scenarios on the \lya\ origin \cite[e.g.][]{Mas-Ribas+17}. 
In fact, differently from the several phenomena behind the \lya\ emission, the nebular UV continuum radiation is only produced in the ISM around star-forming regions via recombinations of hydrogen ions, while the Balmer lines arise via both recombinations and fluorescence. 
Therefore, by comparing Balmer lines and UV continuum SB profiles we can determine the importance of nebular against fluorescent emission, while the comparison between Balmer lines and \lya\ constrains the impact of scattering phenomena.
These studies, however, require deep multi-wavelength observations and are hampered by the faintness and small size (both intrinsic and apparent) of distant galaxies.

One possible solution to solve this puzzle is through the study of strongly lensed LAEs \cite[e.g.][]{Swinbank+07, Karman+15, Caminha+16, Patricio+16, Vanzella+16, Smit+17, Claeyssens+19, Vanzella+20, Chen+21, Iani+21, Claeyssens+22}.
In fact, both lensing effects of magnification and stretching can allow to reach faint fluxes and small scales in short observing time, even though robust lensing models have to be developed in order to correctly interpret and compare the results.

In this context, we study a high-redshift ($z\sim 3.7$) \lya-emitter lensed by the Abell 2895 \cite[hereafter A2895,][]{Abell+58} galaxy cluster ($z\approx0.227$).
Our target (presented in \citealt{Livermore+15} as Abell 2895b) has three multiple images (M1, M2 and M3) located at the celestial coordinates (right ascension, declination) of ($1^h 18^m 11.127^s$, $-26^\circ 57' 59.36''$), ($1^h 18^m 10.543^s$, $-26^\circ 58' 10.56''$), and ($1^h 18^m 10.439^s$, $-26^\circ 58' 14.36''$).
The multiple images are mirrored with respect to the lensing critical line, i.e. the line of infinite magnification.
To investigate the physical properties of our target, we gather a multi-wavelength dataset that covers its rest-frame ultraviolet (UV) and optical emission. 
We study the galaxy rest-frame UV thanks to 
VLT/MUSE optical integral-field spectroscopy with adaptive optics (AO). 
With VLT/SINFONI AO-assisted near-infrared integral-field spectroscopy we probe the rest-frame optical emission of the target.
Thanks to the image multiplicity and the high magnification factor \cite[$\mu=9\pm2$ for the M3 image,][]{Livermore+15}, we are able to probe with unprecedented detail the properties of this source.

This Paper is organised as follows.
In Section~\ref{sec:data}, we present the observations in our hand and on which this paper is based. We also briefly discuss the main steps followed for the data reduction.
In Section~\ref{sec:analysis}, we summarise the technique adopted to obtain the UV and optical spectrum of our target and their analysis. We also describe the procedure to obtain pseudo-NB images of the main emission lines, as well as the UV continuum. 
In Section~\ref{sec:galaxy_prop}, we derive the main properties of our target (e.g. dust extinction, metallicity, star formation rate) based on the analysis of the main spectral features derived from its UV and optical spectra. We also analyse in detail the galaxy \lya\ emission.
In Section~\ref{sec:discussion}, we summarise and discuss our findings.

In this Paper, we adopt a Flat $\Lambda$-CDM cosmology with $\Omega_\Lambda = 0.7$, $\Omega_m = 0.3$, and $H_0 = 70\ \text{km/s/Mpc}$.
All the wavelengths presented in the following (also in the ions nomenclature) are in vacuum.

\section{Observations and Data reduction}
\label{sec:data}
In the following, we describe the observations and the steps performed for the data reduction.
The properties of the 
MUSE observations and the description of the lensing model adopted were already presented in \cite{Iani+21}.
Hence, hereafter we only provide a short description of the MUSE observing programme. We refer the reader to \cite{Iani+21} for further details. 

During the analysis of our target, we serendipitously found additional \lya-emitters at $z\simeq 4.57$, $4.65$ and $4.92$. 
We report their properties in Appendix~\ref{sec:additional_laes}.


\subsection{MUSE data}
\label{subsec:muse_data}
We observed the central region of the A2895 galaxy cluster with VLT/MUSE \citep{Bacon+10}, in Wide Field Mode (WFM, $1'\times 1'$ field-of-view) and with Ground-Layer Adaptive Optics (GLAO) provided by the GALACSI module \citep{Arsenault+08,Strobele+12}. 
The observations were carried out during the 2017 Science Verification of GALACSI (\citealt{leibundgut+17}; Programme ID: 60.A-9195(A), PI: A. Zanella), and in August 2019 (Programme ID: 0102.B-0741(A), PI: A. Zanella), for a total exposure time of 5 hours.
The MUSE WFM observations cover a wavelength range $\Delta \lambda = 4750 - 9350$~\AA\ (nominal) with a spectral resolution $R \sim 3000$.
We reduce the data following the standard reduction procedure (i.e. corrections for bias, flat-field, wavelength and flux calibration, atmospheric extinction and astrometric correction) by means of the ESO reduction pipeline\footnote{\url{https://www.eso.org/sci/software/cpl/esorex.html}} (\textsc{esorex}), version 2.4.1 (\citealt{Weilbacher+12},\citeyear{Weilbacher+14}), and the \textit{Zurich Atmosphere Purge} software (\textsc{zap}\footnote{\url{https://zap.readthedocs.io/en/latest/}} version 2.1, \citealt{Soto+16}) to properly account for sky residuals.
We reconstruct the MUSE PSF via the publicly available algorithm \textsc{psfr}\footnote{\url{https://muse-psfr.readthedocs.io/en/latest/}} \cite[][]{Fusco+20}, and find a FWHM of $0.4''$, as requested for the observations.

\subsection{SINFONI data}
\label{subsec:sinfoni_reduction}
The multiple image M3 of our target was observed with the K-band grating ($\Delta\lambda = 1.95-2.45~\mu{\rm m}$, $R \sim 4000$) of SINFONI \citep{Eisenhauer+03,Bonnet+04}, between August 29th and September 25th 2011 (Programme ID:  085.B-0848(A), PI: J. Richard), for a total exposure time of 6h with AO (natural guide star mode).
The image quality of the observations is $\simeq 0.17''$ in K band, as measured from the standard telluric star observed close in time and airmass to the target, and used for flux calibration.
We reduce the data with the ESO SINFONI pipeline (\textsc{esorex} version 3.13.2) following the standard procedure: correction for dark current, bad pixels and distortions, flat field and wavelength calibration.
We also correct for telluric features, flux calibrate and stack science exposures within the same OB. 
After the reduction of the single OBs, we correct their wavelength calibration for the barycentric velocity, a step that is not automatically performed by the reduction pipeline.
We register the astrometry of the final SINFONI datacube to the one of MUSE by minimising the spatial offset between the centroid of the [OIII]$\lambda5008$ emission and that of the target UV continuum.
We follow this procedure since no other target falls within the SINFONI FoV and the optical continuum of our source is undetected (see Section~\ref{sec:analysis} and bottom panel of Figure~\ref{fig:spectra}). 
Furthermore, the overlap between UV continuum and optical [OIII]$\lambda5008$ emission ensures the spatial overlap with more energetic transitions of the same ion, i.e. OIII]$\lambda\lambda1660,1666$.


\section{Analysis}
\label{sec:analysis}
In the following Section, we report the procedures adopted to extract the UV and optical spectrum of our target and the methodology applied to derive the main properties of the spectral features presented in Table~\ref{tab:table_fitlines} (e.g. line fluxes, equivalent widths).

\subsection{Extraction of the spectra}
\label{subsec:nb-images_spectra}
Following \cite{Iani+21}, we extract the UV and optical spectrum of our target considering the spatial extent of the galaxy brightest UV and optical emission lines, i.e. the \lya~and [OIII]$\lambda5008$, respectively.
As a first step, we create pseudo-narrow band (NB) images that maximise the signal-to-noise ratio (SNR) of these lines \cite[refer to][for more details]{Iani+21}.

We create maps of SNR that we use to define the areas where to extract the galaxy optical and UV spectrum. 
The extraction is performed by summing up the spectra corrected for the lensing magnification factor of all the spaxels where the SNR is $\geq 2.5$.
Since the MUSE data cover all the three multiple images (M1, M2, M3) of our target, while SINFONI observed only M3, we repeat this procedure separately for each multiple image, after cleaning the MUSE UV spectrum from the optical stellar continuum of the A2895 BCG \cite[see][]{Iani+21}.
Finally, we average the UV spectrum of each multiple image to increase the SNR.
In Figure~\ref{fig:spectra} we present the UV (upper panel) and optical (lower panel) spectra of our target.

\subsection{Emission and absorption line measurements}
\label{subsec:lines_fit}
The spectra display several UV an optical emission lines as well as a few weak UV absorption features. 
We fit all the lines (but the \lya) with a Gaussian profile, after modelling their local stellar continuum.
We study the \lya\ separately because of its resonant nature and asymmetric spectral profile, see Section~\ref{subsec:lya_fit}.
To determine the uncertainties on the values derived from the Gaussian fit of each line, we perform 1000 Monte Carlo realisations of the spectra. Each realisation is drawn randomly from a Gaussian distribution with mean and variance corresponding to the observed spectrum flux and variance. 
We define the uncertainty on the line properties as the half distance between the 16th and 84th percentiles.
In Table~\ref{tab:table_fitlines}, we present the results of our fit for all the lines with an estimated SNR $>3$. 
As in \cite{Iani+21}, we add to the final errors also the flux systematic uncertainties due to absolute flux calibration, equal to 5\% and 20\% for MUSE\footnote{We tested the MUSE flux calibration against the \hst/ACS F606W observations (the only available \hst\ image in A2895, SNAP program 10881, PI. G. Smith) for a total of 25 sources falling in the MUSE FoV, and found a good agreement (median magnitudes offset $\leq 0.01$ mag).} and SINFONI data, respectively.

By averaging the wavelength position of the emission lines (but \lya) we estimate the galaxy systemic redshift $z_{\rm sys}=3.72096\pm0.00012$.
For the redshift estimate, we do not consider absorption lines as they are weak spectral features in our UV spectrum and can be blue-shifted whenever the galaxy ISM is characterised by outflows \cite[e.g.][]{Pettini+00,DessaugesZavadsky+10,Patricio+16}.

Finally, we measure the rest-frame equivalent width (EW$_0$) of each line.
We use a definition of EW$_0$ in which negative values indicate emission, while positive values refer to absorption.
Since the optical continuum of the galaxy is not detected in the SINFONI data, we report a 3$\sigma$ upper-limit\footnote{We estimate $\sigma$ as the median of the error spectrum in the wavelength range within which the line fit is performed.} on the line flux and that corresponds to a 3$\sigma$ lower limit on the line EW$_0$.

\afterpage{
\begin{landscape}
\begin{figure}
    \includegraphics[width=1.3\textwidth]{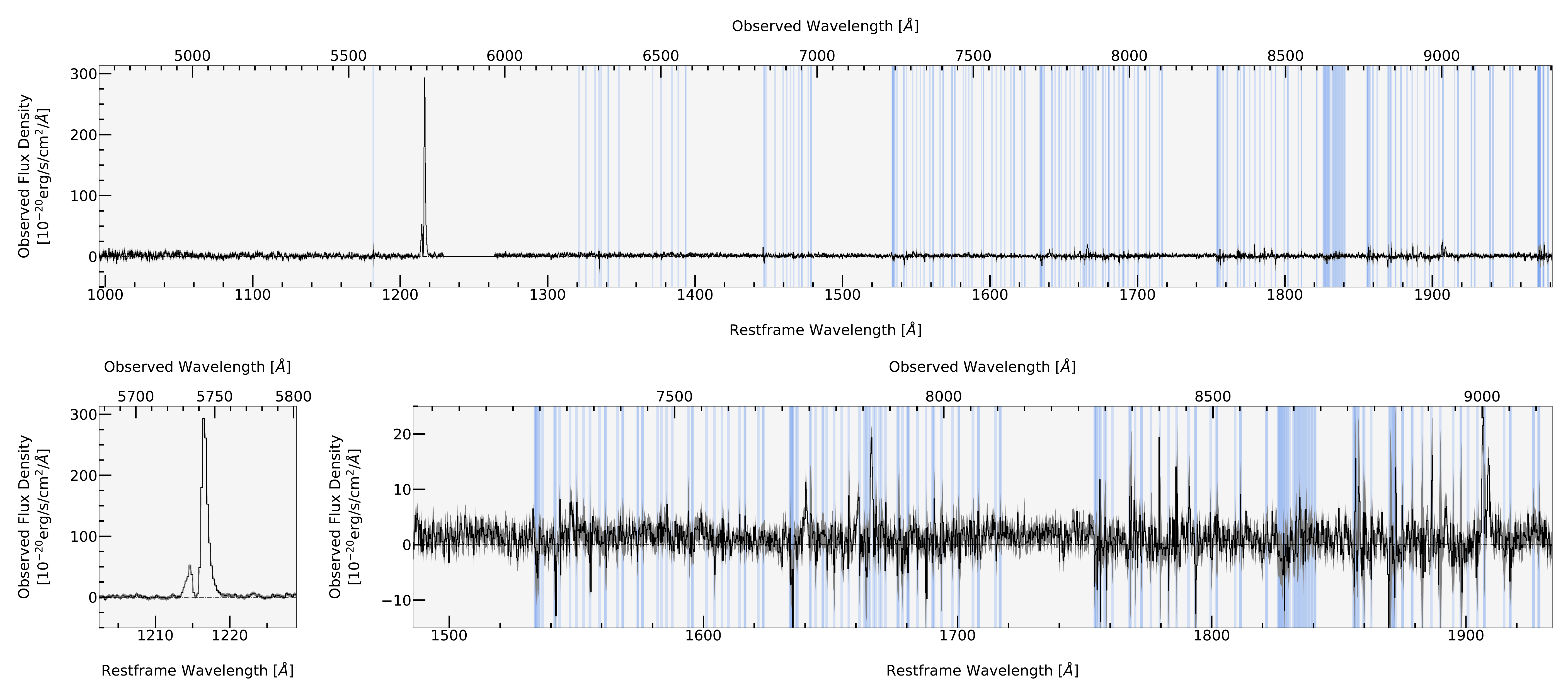}\\
    \includegraphics[width=1.3\textwidth]{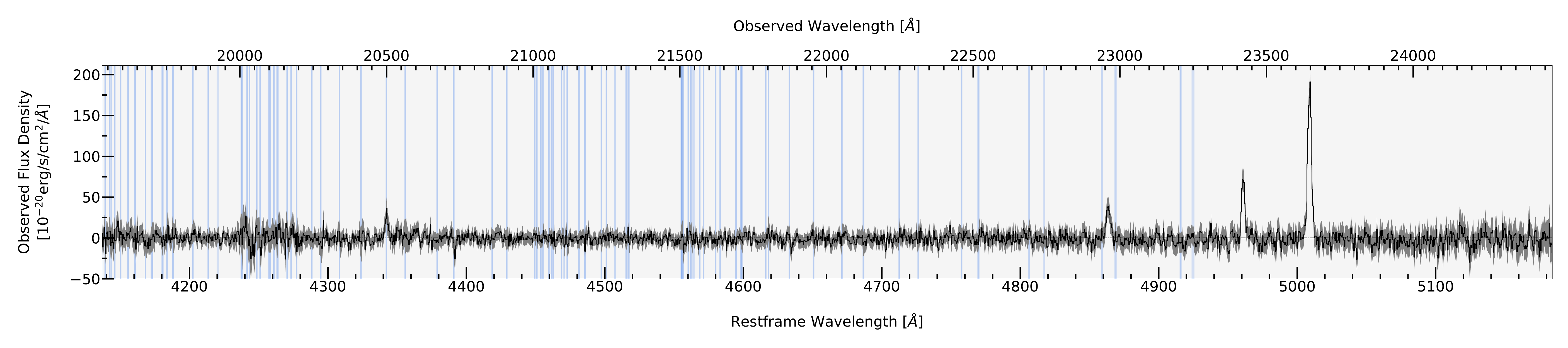}
    \caption{The rest-frame UV (upper panels) and optical spectrum (bottom panel) of our target galaxy. The grey-shaded regions display the $\pm1\sigma$ error around the spectra, while the vertical light-blue solid lines show the wavelength position of strong telluric lines.
    In the central panels we present a zoom-in of the UV spectrum around the \lya\ line (left panel) and the UV continuum redwards the \lya\ and MUSE Notch filter (right panel).}
    \label{fig:spectra}
\end{figure}
\end{landscape}
}

\section{Galaxy properties}
\label{sec:galaxy_prop}
In the following Section, we derive the physical properties (e.g. dust extinction, nebular metallicity, star formation rate) of our target.

\subsection{Dust extinction}
\label{subsec:beta_slope}
We estimate the dust extinction in a twofold way: by considering the slope of the UV continuum ($\beta$-slope) and from the ratio of the Balmer lines \hg/\hb\ (\textit{Balmer decrement}).

For the UV $\beta$-slope, we define 7 spectral windows in the range 1200-2000~\AA\ (see Table~\ref{tab:spectral_windows}) that remove from the fitting procedure all the relevant absorption features of stellar UV spectra, as well as the MUSE Na Notch filter, and fit the observed UV continuum of our target with a power law, i.e. $f(\lambda) \propto \lambda^\beta$ \cite[e.g.][]{Calzetti+94, Castellano+12}.
From the fit we obtain $\beta=-2.6\pm 0.5$. 
Such low value of the $\beta$ parameter is typical of stars with steep blue UV slopes, i.e. young and unobscured stellar populations.
In fact, if we convert the measured $\beta$ into the colour excess of the stellar continuum \ebvcon\ via the relation by \cite{Meurer+99}, we obtain $E(B-V)_{\text{cont}} \lesssim 0.03$. 
Despite the fact that the $\beta$ -- $E(B-V)_{\rm con}$ relation depends on metallicity and star formation history \cite[e.g.][]{Kong+04, Reddy+10, Schaerer+13, Zeimann+15,Reddy+18}, as well as on stellar mass and age \cite[e.g.][]{Buat+12,Zeimann+15, Bouwens+16b}, these values are in line with results found for other low-mass galaxies at intermediate/high-redshifts \cite[e.g.][]{Bouwens+10, Castellano+12, Bouwens+16a, Vanzella+18, Iani+21}. 

Thanks to the simultaneous detection of \hg\ and \hb, we can estimate the nebular extinction due to dust from the observed ratio \hg/\hb.
For our target, we derive (\hg/\hb$)_{\rm obs} = 0.56\pm0.11$.
Adopting the attenuation law by \cite{Calzetti+00} and an intrinsic Balmer decrement of
$(\textnormal{H}\gamma/\textnormal{H}\beta)_{\rm int} = 0.476$ (from case B recombination, e.g. \citealt[][]{Osterbrock+89})
\footnote{Case B recombination assumes that all the ionising photons are processed by the gas ($f_\text{esc}^\text{LyC}=0$). Variations in both the ISM electronic temperature T$_e$ and density n$_e$ affect the expected Balmer ratio \hg$/$\hb. The 0.476 ratio between \hg\ and \hb\ is predicted for T$_e=2\times10^4$K, a n$_e = 10^4$ cm$^{-3}$, and in the absence of an AGN \cite[][]{Osterbrock+89}.}, we obtain a nebular colour excess of $E(B-V)_{\rm neb} = 0.34 ^{+0.41}_{-0.34}$.
Despite the tension between the face-value estimates, the wide errorbars make the nebular estimate compatible with the one obtained from the $\beta$-slope.

In the following we always refer to the observed values of the fluxes without taking into account any dust correction (unless differently specified).

\begin{table*}
    \centering
    \begin{tabular}{lcrrrrrrr}
\hline
\hline
\thead{Line} & \thead{$\lambda_0^a$} & \thead{Flux$^b$} & \thead{EW$_0^c$} & \thead{$z^d$} & \thead{$\sigma^e$} \\
 & \thead{[\AA]} & \thead{[$10^{-20}$ erg/s/cm$^2$]} & \thead{[\AA]} &  & \thead{[km/s]} \\
\hline
NiII       & 1370.132    & -20.2   $\pm$ 5.0      & 1.7    $\pm$ 0.4        & -                         & 104  $\pm$ 45   \\
OV         & 1371.292    & -18.7   $\pm$ 4.1      & 1.6    $\pm$ 0.3        & -                         & 46   $\pm$ 31   \\
SiIV       & 1393.755    & -15.9   $\pm$ 3.8      & 1.4    $\pm$ 0.3        & -                         & 176  $\pm$ 7    \\
SiII       & 1526.707    & -18.4   $\pm$ 4.1      & 2.3    $\pm$ 0.5        & -                         & 42   $\pm$ 41   \\
SiII*      & 1533.431    & 8.0     $\pm$ 1.7      & -1.0   $\pm$ 0.2        & 3.72036 $\pm$ 0.00001     & 6    $\pm$ 1    \\
CIV        & 1548.195    & 29.7    $\pm$ 3.4      & -3.9   $\pm$ 0.4        & 3.72072 $\pm$ 0.00023     & 72   $\pm$ 22   \\
HeII       & 1640.417    & 44.5    $\pm$ 3.5      & -7.8   $\pm$ 0.6        & 3.72087 $\pm$ 0.00008     & 64   $\pm$ 13   \\
OIII{]}    & 1660.809    & 27.4    $\pm$ 4.0      & -4.8   $\pm$ 0.7        & 3.72097 $\pm$ 0.00019     & 34   $\pm$ 14   \\
OIII{]}    & 1666.150    & 79.7    $\pm$ 8.4      & -12.3  $\pm$ 1.3        & 3.72095 $\pm$ 0.00015     & 51   $\pm$ 11   \\
AlII       & 1670.787    & 9.2     $\pm$ 1.4      & -1.4   $\pm$ 0.2        & 3.72041 $\pm$ 0.00001     & 47   $^{+51}_{-47}$   \\
SiIII{]}   & 1892.029    & 25.6    $\pm$ 1.3      & -4.8   $\pm$ 0.3        & 3.72110 $\pm$ 0.00002     & 47   $\pm$ 2    \\
{[}CIII{]} & 1906.680    & 96.6    $\pm$ 11.6     & -18.1  $\pm$ 2.3        & 3.72095 $\pm$ 0.00015     & 44   $\pm$ 17   \\
CIII{]}    & 1908.734    & 72.2    $\pm$ 12.0     & -13.4  $\pm$ 2.3        & 3.72119 $\pm$ 0.00039     & 58   $\pm$ 46   \\
H$\gamma$  & 4341.680    & 365.8   $\pm$ 99.4     & $\leq$-2.3$^\dagger$    & 3.72151 $\pm$ 0.00030     & 14   $^{+162}_{-14}$  \\
{[}OIII{]} & 4364.436    & 178.1   $\pm$ 63.1     & $\leq$-1.2$^\dagger$    & 3.72027 $\pm$ 0.00084     & 67   $\pm$ 53   \\
H$\beta$   & 4862.680    & 652.9   $\pm$ 141.2    & $\leq$-3.9$^\dagger$    & 3.72160 $\pm$ 0.00024     & 62   $\pm$ 28   \\
{[}OIII{]} & 4960.295    & 988.6   $\pm$ 204.9    & $\leq$-5.5$^\dagger$    & 3.72147 $\pm$ 0.00009     & 21   $^{+24}_{-21}$   \\
{[}OIII{]} & 5008.240    & 2909.7  $\pm$ 584.3    & $\leq$-15.5$^\dagger$   & 3.72147 $\pm$ 0.00004     & 50   $\pm$ 5    \\
\hline
    \end{tabular}
    \caption{Properties of the emission and absorption lines with measured SNR $>3$. 
    Unless differently stated, the measurements reported refer to the intrinsic values, i.e. corrected for lensing magnification. 
    $^a$: the wavelengths reported are in vacuum. 
    $^b$: the flux uncertainties have been increased by 5\% (MUSE) and 20\% (SINFONI) to take into account errors on the absolute calibration of the datasets. 
    $^c$: rest-frame EW of the line (the $^\dagger$ highlights lines for which the EW$_0$ has been estimated taking into account an upper limit on the stellar continuum flux). We follow the convention for which the EW of emission lines is reported with negative values (see Section~\ref{subsec:lines_fit}).
    $^d$: estimated redshift of the target according to the wavelength of the best-fit Gaussian peak (only for nebular emission lines, as absorption lines might have blue-shifted spectral profiles due to outflows, see Section \ref{sec:galaxy_prop}). 
    $^e$: velocity dispersion $\sigma$ corrected for instrumental broadening \protect\cite[see][]{Iani+21} and in units of km/s.}
    \label{tab:table_fitlines}
\end{table*}

\begin{table}
    \centering
    \begin{tabular}{ccc}
    \hline
    \hline
    Window  & Wavelength Range \\
    Number  & [\AA]\\
    \hline
    1     & 1268 - 1284\\
    2     & 1360 - 1371\\
    3     & 1407 - 1515\\
    4     & 1562 - 1583\\
    5     & 1677 - 1725\\
    6     & 1760 - 1833\\
    7     & 1930 - 1950\\
    \hline
    \end{tabular}
    \caption{Rest-frame UV spectral windows employed for the measurement of the stellar continuum $\beta$-slope, see Section~\ref{subsec:beta_slope}.}
    \label{tab:spectral_windows}
\end{table}

\subsection{Electron temperature and density, nebular metallicity, ionisation parameter and (C/O) abundance}
\label{subsec:te_ne}
Thanks to the presence of the [OIII] optical lines at 4364, 4959 and 5008\AA, we evaluate the electron temperature T$_e$ of the emitting gas following
the empirical equation by \cite{Proxauf+14}. 
The electron temperature we derive is of T$_e=(2.2\pm0.2)\times10^4$~K.

In a similar manner, because of the detection of the [CIII] doublet at 1908\AA, we determine the electron density $n_e$ from the intensity ratio 
$\textnormal{[CIII]}\lambda1907/\textnormal{CIII]}\lambda1909$.
By means of \textsc{PyNeb} and assuming  T$_e=(2.2\pm0.2)\times10^4$~K, we derive an upper-limit on the $n_e\lesssim 3\cdot 10^4\ {\rm cm}^{-3}$.

\begin{figure}
    \centering
    \includegraphics[width=.98\columnwidth,trim={0cm 0cm 0cm 1cm}]{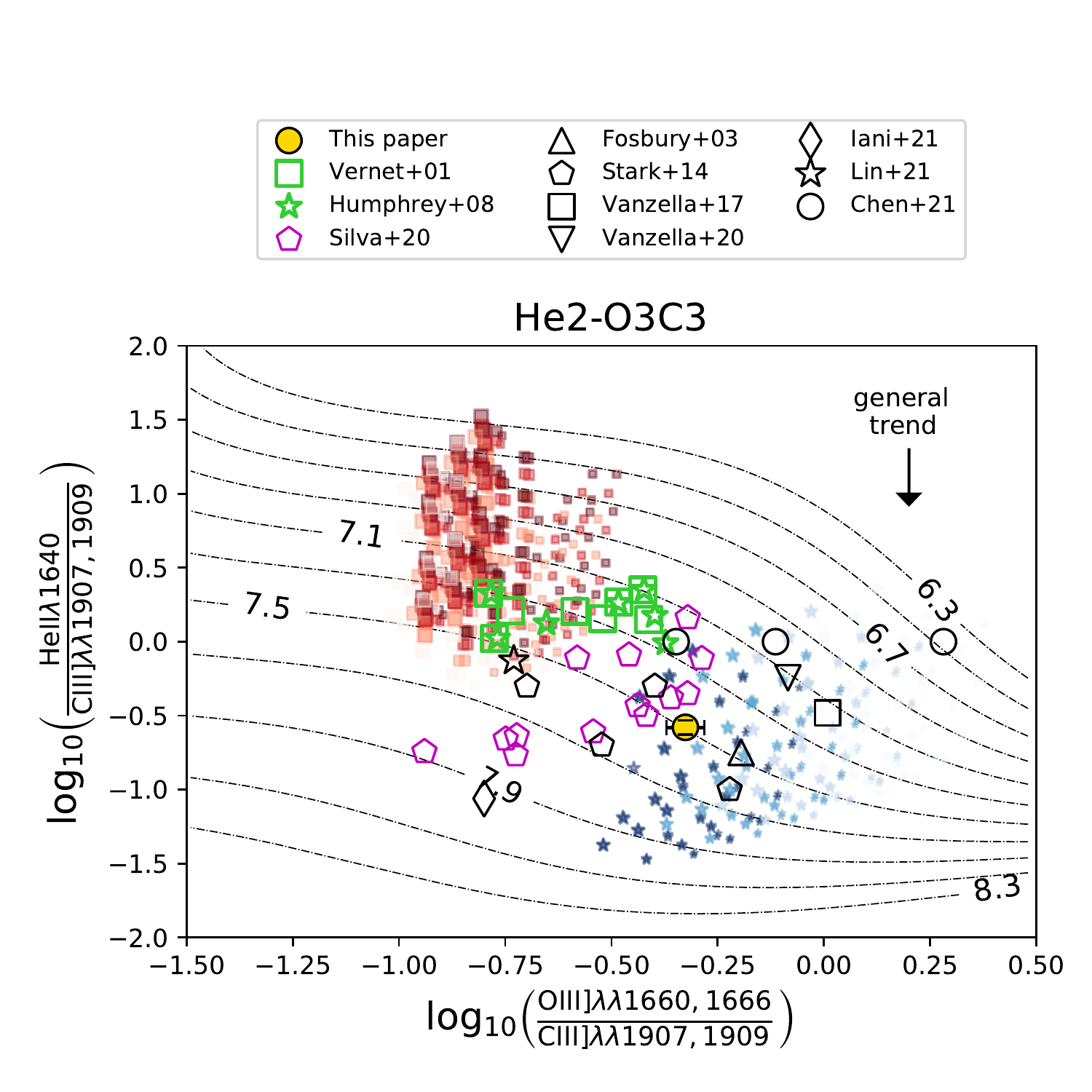}
    \caption{He2-O3C3 diagnostic diagram. The coloured solid lines \protect\cite[from Equation 8 in][]{Byler+20} show how the line ratios vary as a function of the nebular metallicity. The yellow circle displays the position of our target in this diagram, while the open marks are representative points of a sample of intermediate/high-$z$ sources taken from literature \protect\cite[in black,][]{Fosbury+03, Stark+14, Vanzella+17, Vanzella+20, Chen+21, Lin+21, Iani+21}, a sample of high-$z$ radio galaxies \protect\cite[in green,][]{Vernet+01,Humphrey+08} and a sample of type-II quasars \protect\cite[in magenta,][]{Silva+20}. The blue stars show the area of the diagram populated by the stellar models by \protect\cite{Gutkin+16} with sub-solar metallicities  (more details in Section~\ref{subsubsec:diagnostic_diagrams}). The size of these representative points is linked to their ionisation parameter U (larger the size, higher the ionisation parameter), while their shade of blue depends on the (C/O)/(C/O)$_\odot$ (darker the colour, higher the ratio). In a similar way, we present the ratio of nebular emissions for the AGN models by \protect\cite{Feltre+16} with red squares. The size of these representative points is still linked to their ionisation parameter U, while their shade of red depends on the (C/O)/(C/O)$_\odot$ (darker the colour, higher the ratio). Finally, we highlight with a black arrow in the top right-hand side corner the effect of a correction for stellar emission on the HeII$\lambda1640$ line.}
    \label{fig:metallicity}
\end{figure}

The UV emission lines detected in our target spectrum allow us to estimate its nebular metallicity by means of the He2 -- O3C3 diagnostic diagram by \cite{Byler+20}, see Figure~\ref{fig:metallicity}.
The He2 -- O3C3 diagram has been found to robustly determine the ISM metallicity of metal-poor (sub-solar) systems.
Specifically, through Equation 8 by \cite{Byler+20}, we derive an ISM metallicity $12+\log_{10}(O/H)=7.36\pm0.02$.
Assuming a solar value of $12+\log_{10}(O/H)_\odot=8.69\pm0.05$ \citep{APrieto+01},
the derived estimate corresponds to a sub-solar metallicity of ${\rm Z}_{\rm neb} = 0.05\pm 0.02$~Z$_\odot$, i.e. ${\rm Z}_{\rm neb} = (7 \pm 3)\times 10^{-4}$.

In their work, \cite{Byler+20} do not provide any relation to determine the ionisation parameter (U) as in the case of the nebular metallicity. However, from the comparison with the model grids, we estimate $\log_{10}(\text{U})\sim-2.5$. 

We finally investigate the (C/O) abundance of our target by using Equations 6 and 7 by \cite{Perez-Montero+17}.
From the equations, we obtain $\log_{10}({\rm C/O}) = -0.99 \pm 0.23$. If we compare this estimate to the solar value ${\rm (C/O)}_{\odot} = 0.44$ \cite[][]{Gutkin+16}, we derive a ratio ${\rm (C/O)/(C/O)}_{\odot} \simeq 0.23$.
The (C/O) estimate is in agreement with the (C/O) -- metallicity relation and with the values found in metal-poor high-ionisation dwarf local galaxies and halo stars \cite[e.g.][]{Berg+16, Berg+19}. Interestingly, because carbon and oxygen are thought to originate primarily from stars of different mass ranges (with O synthesised mostly in massive stars with ${\rm M} > 10~{\rm M}_\odot$, while C is produced in both massive and intermediate mass stars, i.e. $2 ~{\rm M}_\odot < {\rm M} < 8~{\rm M}_\odot$), the low (C/O) measured could be interpreted as the consequence of a top-heavy IMF and/or the sign that we are looking at a very recent burst of star formation. In fact, while in the first case an overabundance of massive stars could bring to an enhanced production of O over C, the same effect could be obtained if the stellar population had just recently formed and only type II supernovae (SNe) had time to enrich the ISM \cite[$\lesssim 40$~Myr, e.g.][]{Veilleux+05}.

\subsection{The source of the ISM ionisation: AGN, SF, or shocks?}
\label{subsec:agn_or_sfr}
Different mechanisms (e.g. star formation, AGN, shocks) can ionise the ISM of galaxies, thus driving the emission of lines in the UV and optical.
To discriminate between the different processes, several tracers can be probed, e.g. the width of emission lines, the presence of asymmetries and broadening in their shapes, the detection of specific atomic transitions of heavy elements, emission line ratios.

\subsubsection{Detection, shape and width of UV metal lines}
From the UV and optical spectrum, the emission line profiles do not show the presence of blue/red wings nor broad components.
All the emission lines are narrow, having a $\sigma \leq 100$~km/s (see Table~\ref{tab:table_fitlines}) and being unresolved or marginally resolved at the spectral resolution of our data. 
These results disfavour the hypothesis that our target hosts an unobscured (type-I) AGN \citep[e.g.][]{McCarthy+93,Corbin+96,Humphrey+08,Matsuoka+09}.
This is also supported by the fact that the publicly available X-ray catalogue based on \textit{Chandra} observations (Chandra Source Catalog, v. 2.0, \citealt{Evans+19, Evans+20})\footnote{The Chandra Source Catalog (CSC) is available at \url{https://cxc.cfa.harvard.edu/csc2/}.} for this cosmological field does not report any X-ray emitter at our target coordinates nor in its closest vicinity\footnote{The \textit{Chandra} observations for this field (ID proposal: 9429, P.I. G. P. Smith) were carried out with the imaging mode of the Advanced CCD Imaging Spectrometer (ACIS-I) and have a $3\sigma$ depth at the positions of the source of $\simeq 2.3\times10^{-15}\rm erg/s/cm^{-2}$. The {\it Chandra} images are available at \url{https://cda.harvard.edu/chaser/}.}.   

However, we cannot fully rule out the possible presence of an obscured (type-II) AGN. In fact, the UV spectrum of our target shows several high ionisation potential transitions of He and metals, e.g. CIV$\lambda1550$ and HeII$\lambda1640$, whose presence could be explained by AGN activity. 
This hypothesis could also be supported by the high EW$_0$ measured for the HeII line ($\sim -8$\AA). In fact, recent literature has shown that photo-ionisation models fail to reproduce such rest-frame EW with the only contribution of stars \cite[e.g.][]{Berg+18, Nanayakkara+19}.
A line that is generally considered to be the smoking-gun proof of an AGN is NV$\lambda1240$, having an ionisation potential of $\sim 78$ eV, difficult to explain with the typical emission of stellar populations \cite[e.g.][]{Hainline+11, Laporte+17, Grazian+20}.
Yet, the MUSE Na Notch filter prevents us from detecting this line.

\subsubsection{UV spectroscopic diagnostic diagrams}
\label{subsubsec:diagnostic_diagrams}
A possible way to better constrain the nature of our target is to consider empirical spectroscopic diagnostic diagrams based on the ratio of UV lines. 
Similarly to the optical BPT and BPT-like diagrams (\citealt{Baldwin+81, Veilleux+87, Dopita+95}), the ratio between close-by UV emission lines is effective in discriminating among different mechanisms of ISM ionisation \cite[e.g.][]{Feltre+16, Nakajima+18, Hirschmann+19}. 
Besides, the fact that these diagrams feature the intensity ratio of lines that are close in wavelength makes them rather insensitive to dust extinction.  

\begin{figure*}
    \centering
    \includegraphics[width=.48\textwidth,trim={0cm 0cm 0cm 1cm}]{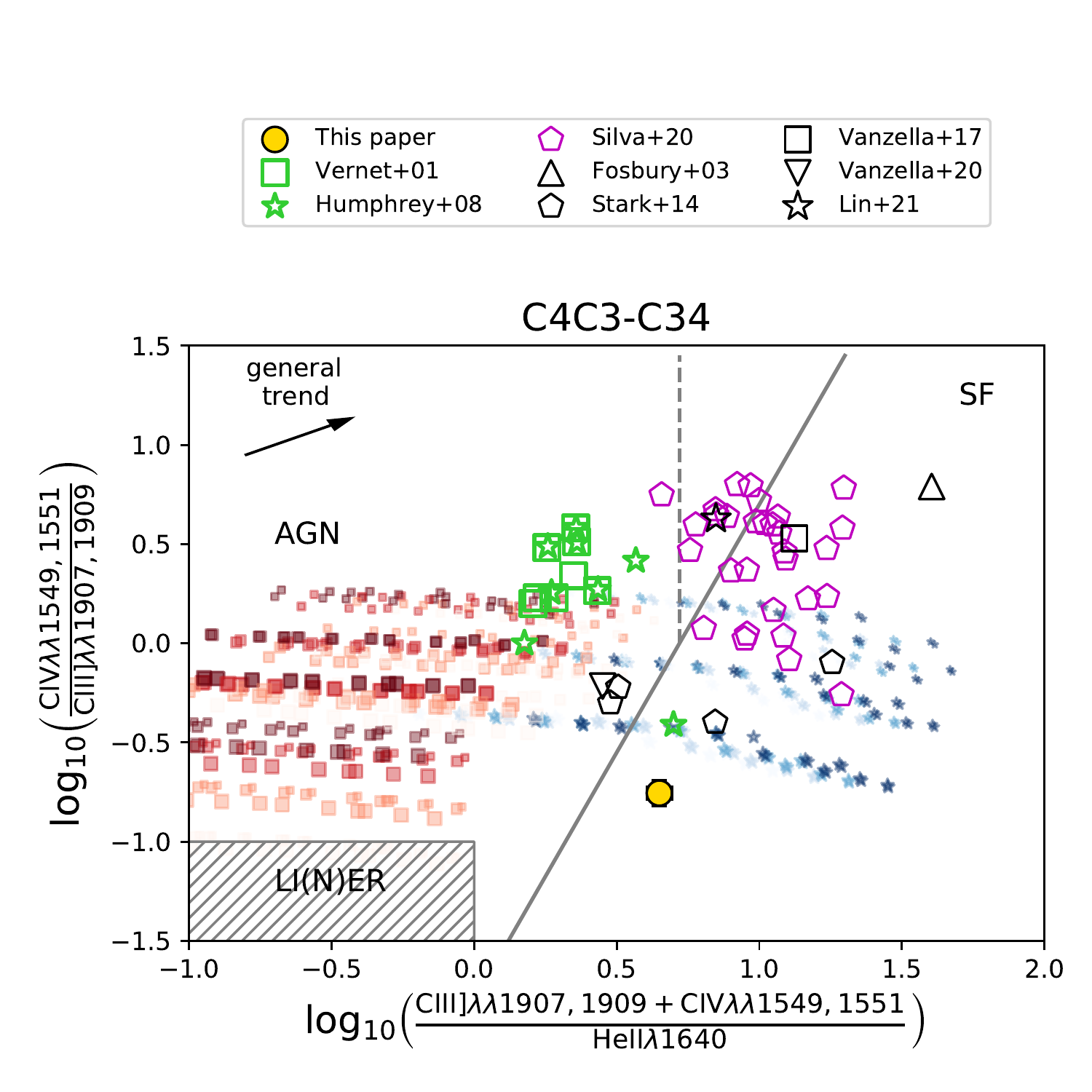}
    \
    \includegraphics[width=.48\textwidth,trim={0cm 0cm 0cm 1cm}]{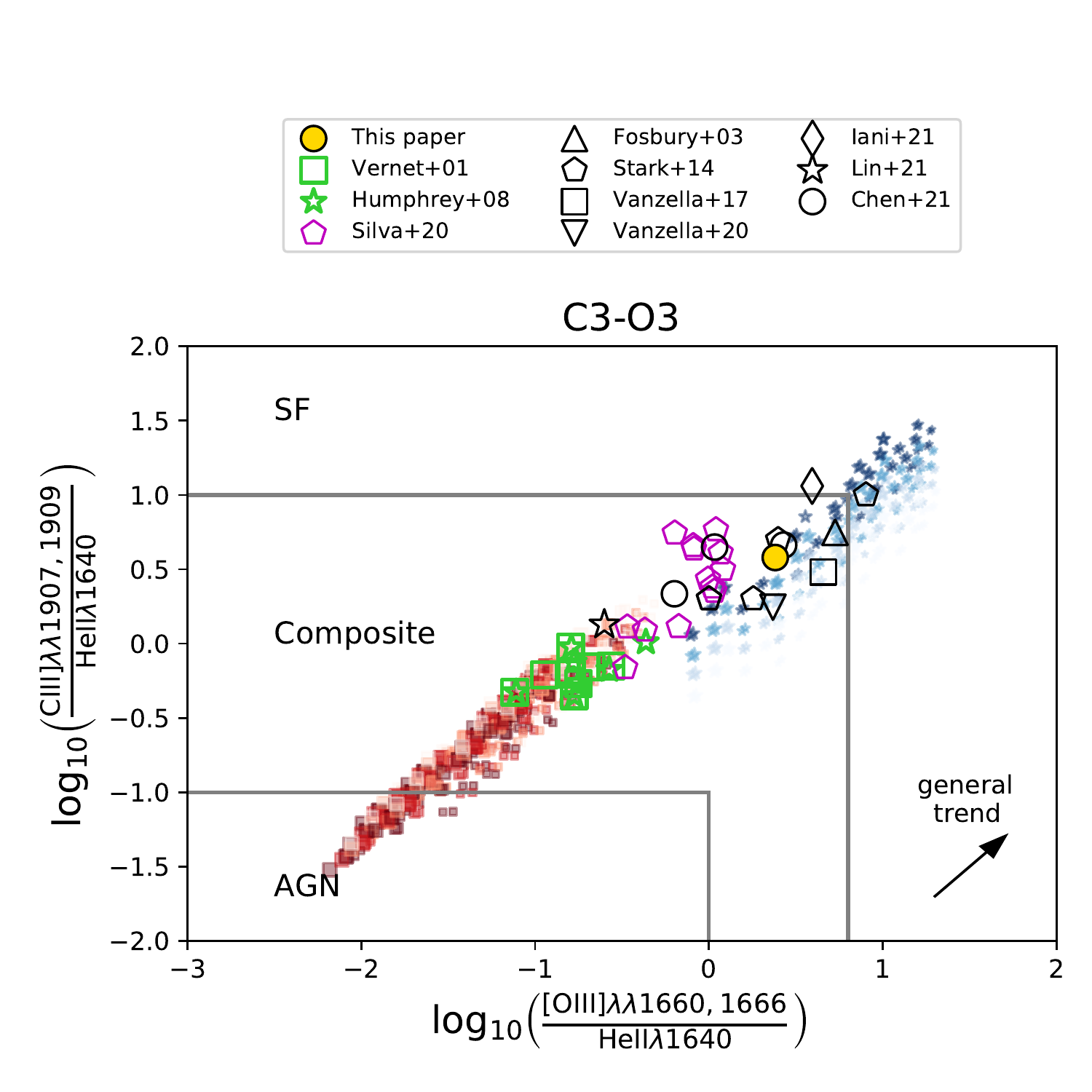}
    \caption{C4C3-C34 (left panel) and C3-O3 (right panel) empirical diagnostic diagrams. The yellow circle displays the position of our target in these diagrams, while the open marks display the position of a sample of intermediate/high-$z$ sources taken from literature \protect\cite[in black,][]{Fosbury+03, Stark+14, Vanzella+17, Vanzella+20, Chen+21, Lin+21, Iani+21}, a sample of high-$z$ radio galaxies \protect\cite[in green,][]{Vernet+01,Humphrey+08} and a sample of type-II quasars \protect\cite[in magenta,][]{Silva+20}. The blue stars show the area of the diagram populated by the stellar models by \protect\cite{Gutkin+16}. In a similar way, we present the ratio of nebular emissions for the AGN models by \protect\cite{Feltre+16} with red squares. For both models, the size and colour of the representative points follow the convention presented in Figure~\ref{fig:metallicity}. In both panels, the black arrow in the corner highlight the effect of a correction of the HeII$\lambda1640$ and CIV$\lambda1550$ (only in the C4C3-C34 panel) emission. In the C4C3-C34 diagram, the demarcation lines are taken from \protect\cite{Nakajima+18}. In the C3-O3 diagram we resort to the separation lines by \protect\cite{Hirschmann+19}.}
    \label{fig:diagnostics}
\end{figure*}

In our work, we exploit the C4C3-C34, C3-O3 and He2-O3C3 diagrams \cite[e.g.][]{Nakajima+18,Hirschmann+19, Byler+20}, see Figures~\ref{fig:metallicity},\ref{fig:diagnostics}. 
For the C4C3-C34 diagram, we highlight the demarcation lines between AGN and star-forming galaxies presented in \cite{Nakajima+18}. In the case of the C3-O3 diagnostic, we resort to those of \cite{Hirschmann+19}.
We populate the panels with other metal-poor intermediate/high-redshift sources taken from the literature \citep{Fosbury+03,Stark+14,Vanzella+17,Vanzella+20,Chen+21,Lin+21,Iani+21}, a sample of high-$z$ radio galaxies (HzRG) by \cite{Vernet+01} and \cite{Humphrey+08}, a sample of type-II quasars (QSO2s) from \cite{Silva+20}, as well as from theoretical models\footnote{The models are available at \url{http://www.iap.fr/neogal/models.html}.} of both star-forming galaxies \cite[blue stars,][]{Gutkin+16} and AGNs \cite[red squares,][]{Feltre+16}. 
We limit both theoretical models to sub-solar metallicities ${\rm Z} = [0.0001,0.0002,0.0005,0.001]$ and ionisation parameters $\log_{10}(\text{U})=[-3,-2.5,-2]$. 
As for the stellar models\footnote{The line fluxes presented in the stellar models by \cite{Gutkin+16} are in units of solar bolometric luminosity (i.e. $3.826\cdot10^{33}\ {\rm erg}/{\rm s}$) per unit SFR (in ${\rm M}_\odot/{\rm yr}$), and assuming a constant SFR with a Chabrier IMF sustained for $10^8$ yr.}, we apply an additional cut to ${\rm (C/O)}/{\rm (C/O)}_\odot = [0.14,0.20,0.27,0.38]$ and an upper cut-off stellar mass for the IMF $M_{\rm up} = 300\ {\rm M}_\odot$. 
The constraints on the ISM metallicity, ionisation parameter and ${\rm (C/O)}$ abundance are based on the estimates that we retrieved in Section~\ref{subsec:te_ne}.
On the contrary, we do not limit the stellar and AGN models neither in gas density $n_{\rm H} = [100,1000]\ {\rm cm}^{-3}$ nor in dust-to-metal mass ratios $\xi_d = [0.1,0.3,0.5]$. 
Finally, the AGN models\footnote{The line fluxes presented in the AGN models by \cite{Feltre+16} correspond to an accretion luminosity of the central source of $10^{45}$ erg$/$s, and that line luminosities scale linearly with this quantity.} have UV spectral slopes $\alpha$ ($f_\nu \propto \nu^\alpha$) ranging from $[-2,-1.7,-1.4,-1.2]$.

On one hand, according to the demarcation lines by \cite{Nakajima+18} (grey lines in the C4C3-C34 diagram), our target seems to be a star forming galaxy. 
On the other hand, based on the redshift-independent separation criteria by \cite{Hirschmann+19}, the representative point of our target in the C3-O3 diagram lays at the edge of the {\it composite} region, i.e. galaxies that are characterised by both on-going star formation and AGN activity.
Depending on the luminosity of the hosted AGN $L_{\rm AGN}$ with respect to star formation ($f = L_{\rm AGN}/L_{\rm SF}$), the representative points of sources in the composite region can tend more towards the SF ($f < 0.5$) or AGN ($f > 0.5$) areas.
According to the position of our galaxy, the contribution of the AGN luminosity would be less than 0.5 times the one of SF. 
The composite locus is, however, an area of the diagnostic diagram that is not uniquely determined: theoretical models show that it can also be consistently populated by galaxies whose emission lines originate from shocks (the so-called {\it shock-dominated} galaxies, e.g.  \citealt{Hirschmann+19}), as well as \textit{pure} SF and AGN models \cite[][]{Gutkin+16, Feltre+16}. 
Hence, it is difficult to pinpoint the mechanism that originates the observed UV lines. 
Nonetheless, by comparing the position of our target with the available theoretical models, we observe a higher compatibility of our estimates with the synthetic stellar models by \cite{Gutkin+16}.
This is also observed in the He2-O3C3 diagram where we do not have separation lines for the different mechanisms. However, we record a net separation between the AGN and star formation models, with the representative point of our galaxy laying in the area populated by the models by \cite{Gutkin+16}.

A word of caution has to be spent, however, in interpreting the results presented in these diagnostics. 
In fact, the diagrams should be constructed considering the ratio of nebular lines only, i.e. lines emitted by the ionised ISM.
However, the intensity of CIV can be affected by stellar absorption whereas the HeII emission can come from both stars and the ionised ISM \citep{Brinchmann+08, Erb+10}.
This implies that both the CIV and HeII observed intensities should be corrected for these effects. The corrections would bring to an enhancement of the CIV flux and a decrease of the HeII intensity. 
In the pure AGN scenario there would be no need of such corrections for HeII \cite[see][and references therein]{Nakajima+18}. 
The CIV and HeII intensity used to construct the diagrams presented here have been directly obtained from the fit of the observed spectrum of our target. 
We are therefore assuming that the HeII flux is only originated by the ionised ISM, while the CIV is not affected by stellar absorption and by the possible presence of a P-Cygni profile originated by stellar winds.
With the data at our disposal, we cannot apply any realistic flux correction to CIV and HeII. 
Besides, the corrections have been found to depend on several stellar population parameters as metallicity and age, as well as if binary stellar evolution is taken into account \citep{Erb+10, Steidel+16}. 
For completeness, however, we report with an arrow the general trend of the above-mentioned corrections in the corner of each panel (the length of the arrow along the two axis is not related to the relative strength of the correction).  
We highlight how, for all diagrams, the introduction of corrections in the CIV and HeII flux would move the representative point of our target even more towards the region of star-forming galaxies.
Besides, we observe a net separation between our target and the population of HzRGs from \cite{Vernet+01}
and \cite{Humphrey+08}, as well as the sample of type-II quasars from \cite{Silva+20}.
This result reinforces the idea that our target is a star-forming galaxy.

\subsection{Star formation rate}
\label{subsec:sfr}
Assuming that our target is a star forming galaxy, we estimate its integrated star formation rate (SFR) in a two-fold way: from the luminosity of the Hydrogen Balmer lines, and from the luminosity of the stellar UV continuum.
In both cases, we start from the recipes by \cite{Kennicutt+98} modified for a Chabrier IMF\footnote{To transform from a Salpeter to a Chabrier IMF, we divide by a 1.7 factor.}. 

For the Balmer lines, the original \cite{Kennicutt+98} prescription is based on the direct conversion of the \ha\ luminosity into SFR. 
The estimate that can be derived through this method allows us to probe the so-called {\it instantaneous} star formation, i.e. the galaxy star formation activity over its last 10 Myr. 
However, the \ha\ line is out of the wavelength coverage of our data. 
To overcome this problem, we convert the \hb\ flux into \ha\ assuming case B recombination \citep{Osterbrock+06}\footnote{According to case B recombination the intrinsic ratio \ha/\hb\ is equal to $2.74$ for a T$_e=2\times 10^4$~K and $n_e=10^4~{\rm cm}^{-3}$}.
We derive a SFR $= 10.7\pm2.3~\text{M}_\odot/\text{yr}$. 

Similarly, we derive the SFR from the rest-frame UV stellar continuum luminosity. Compared to the \ha\ -- SFR conversion, the UV -- SFR relation is based on stronger assumptions, among which a continuous and well-behaved star formation history, ongoing typically for at least $100$ Myr. 
Following \cite{Kennicutt+98}, we convert the galaxy rest-frame luminosity at 1500\AA,\ $L_\nu(1500\text{\AA})$. 
We extrapolate $L_\nu(1500\text{\AA})$ from the fit of the UV continuum with a power-law, see Section~\ref{subsec:beta_slope}, and obtain $L_\nu(1500\text{\AA}) = (7.63\pm 0.41)\cdot 10^{27}\ {\rm erg/s/Hz}$ and a SFR $= 0.64\pm0.03~\text{M}_\odot/\text{yr}$.
This estimate, however, does not take into account the correction for the contribution of the nebular UV continuum emission.
Although typically weaker at the UV wavelengths than in the optical and near-infrared, the nebular continuum originates from free-free, free-bound and {\it two-photons}\footnote{The two-photons continuum is a bound-bound process where the excited $2s$ state of Hydrogen decays to the $1s$ state via the simultaneous emission of two photons. The energy of the two photons produces a bump in the UV spectrum at $\approx 1500$\AA\ \cite[][]{Byler+17}.} transitions, and its contribution to the overall continuum emission depends on several physical parameters among which the ionisation parameter ${\rm U}$, the temperature of the emitting HII region, the nebular metallicity, the age of the stellar population emitting the ionising photons \cite[][]{Byler+17}.  
By means of \textsc{PyNeb} \cite[][]{Luridiana+15} and following \cite{Fernandez+18}\footnote{To estimate the monochromatic flux of the nebular continuum, we apply equation 15 in \cite{Fernandez+18}:
\begin{equation*}
F_{\rm neb} \simeq \frac{\gamma}{\alpha_{{\rm H}\beta}^{\rm eff}\cdot h\nu_{{\rm H}_\beta}}\cdot F({\rm H}\beta) 
\end{equation*}
where $\gamma$ is the nebular continuum emissivity (in erg$\cdot$cm$^3$/s/\AA) as estimated from \textsc{PyNeb} \cite[][]{Luridiana+15} assuming $T_e=2.2\times 10^4$K and $n_e=3\times10^4 {\rm cm}^{-3}$ (see Section~\ref{subsec:te_ne}), $\alpha_{{\rm H}\beta}^{\rm eff}$ is the \hb\ effective recombination coefficient \cite[$1.61\cdot 10^{-14}$ cm$^3$/s,][p. 84]{Osterbrock+89}, $h\nu_{{\rm H}\beta}$ is the energy associated to the \hb\ transition (i.e. $4.08\cdot10^{-12}$ erg) and $F({\rm H}\beta)$ is the measured \hb\ flux (in erg/s/cm$^2$).}, we estimate a contribution of the nebular continuum to the observed flux at 1500\AA\ of 50-60 percent, with the precise estimate depending on the HeI/HI and HeII/HI abundances. 
If we assume T$_e = 2\times 10^4$~K, $n_e=10^4~{\rm cm}^{-3}$, ${\rm HeI/HI} = 0.08$ and ${\rm HeII/HI} = 0.02$, we derive $f(1500{\rm \AA})_{\rm neb}/f(1500{\rm \AA})_{\rm obs} = 0.53$. 
Hence, correcting the observed UV luminosity at 1500~\AA\ for nebular emission, would lower the previous SFR estimate to $0.34\pm0.02~{\rm M}_\odot/{\rm yr}$.

According to our measurements, the
$\text{SFR(\ha)}$ is about 18 times higher (28$\times$ after correcting for the nebular emission) than the $\text{SFR(1500\AA)}$. 
This significant discrepancy is possibly due to the fact that the assumptions behind the conversion of the two tracers are not fully fulfilled. 
In particular, the conversion factor in the UV -- SFR relation is known to be severely underestimated in the case of a young stellar population \citep[$\lesssim 10$~Myr, e.g.][]{Kennicutt+98, Calzetti+13}.
In this regard, the discrepancy between the SFR inferred from the Balmer lines and the UV could favour a \textit{bursty} on-going star-formation against a steady state process scenario \cite[e.g.][]{Guo+16, Faisst+19, Atek+22}. 
Additionally, deviations on the shape of the IMF and its mass limits, the stellar metallicity and the number of ionising photons produced can play an important role.
To estimate the correction factor that should be taken into account when converting the UV luminosity into SFR in the case of a single burst star-formation history, we resort to the \textsc{bpass} models\footnote{\url{https://bpass.auckland.ac.nz/9.html}} \cite[v. 2.2.1][]{Eldridge+17, Stanway+18}. 
If we assume both a Chabrier and top-heavy IMF\footnote{The available \textsc{bpass} top-heavy IMF is given by:
\begin{equation*}
    N(M<M_{\rm max}) \propto \int_{0.1}^{M_1} \left(\frac{M}{{\rm M}_\odot}\right)^{\alpha_1}dM + M_1^{\alpha_1}\int_{M_1}^{M_2} \left(\frac{M}{{\rm M}_\odot}\right)^{\alpha_2}dM 
\end{equation*}
where $M_1 = 0.5 {\rm M}_\odot$, $M_2 = 300 {\rm M}_\odot$, $\alpha_1 = -1.3$ and $\alpha_2=-2.0$. 
For more details on the \textsc{bpass} IMFs, we refer the reader to the \textsc{bpass} user manual.} with an upper cut-off mass of $300 ~{\rm M}_\odot$ and a sub-solar metallicity ($10^{-5}\leq {\rm Z} \leq 10^{-3}$), we obtain a correction factor to the L(UV) -- SFR relation by \cite{Kennicutt+98} of about 8 at 2-3 Myr, see Figure~\ref{fig:age_tracks}. 
The magnitude of the correction decreases rapidly with time and shrinks down to a factor of about 2 at 10 Myr.
In particular, the \textsc{bpass} tracks show that both more top-heavy IMFs and lower stellar metallicities increase the magnitude of the correction factor.
A similar result is also recovered if we consider spectro-photmetric synthetic models derived via the \textsc{Starburst99}\footnote{\url{https://www.stsci.edu/science/starburst99/docs/parameters.html}} code \cite[][]{Leithere+99}.
In the case of \textsc{Starburst99}, we construct the models taking into account a single burst star formation history (SFH), a Chabrier-like and top-heavy IMF\footnote{For the Chabrier-like IMF, we assume an exponent of -1.3 between 0.01 M$_\odot$ and 0.5 M$_\odot$, and -2.3 for stellar masses in the interval 0.5 - 300 M$_\odot$. For the top-heavy IMF, we adopt -1.35 between 0.01 M$_\odot$ and 300 M$_\odot$ \cite[e.g.][]{Zanella+15}.}, the Padova stellar tracks 
and a sub-solar metallicity ${\rm Z} = 4\times 10^{-4}$.
For the \textsc{Starburst99} tracks, the correction factor is smaller, $\sim 4$ at 2-3 Myr \cite[in agreement with the estimates reported by][]{Santini+14}, if compared to \textsc{bpass}.
This result clearly highlights the impact that the introduction of binary stellar systems (as in \textsc{BPASS}) has in the modelling of the properties of stellar populations in galaxies \cite[see also][]{Reddy+22}.

Despite significant, an increase of a factor 8 in the estimate of the SFR(1500\AA) is not able to explain alone the discrepancy between our two SFR estimates.
The additional introduction of dust extinction could alleviate the tension, see Figure~\ref{fig:age_tracks}.

\subsection{Galaxy mass and stellar population age}
\label{subsec:mass_age}
The absence of stellar continuum detection in SINFONI and of additional rest-frame optical and near-infrared (NIR) data prevents us from deriving a robust estimate of the stellar mass $M_\star$ for our target.
However{, \bf despite being an indirect and uncertain method}, we can infer an upper limit on $M_\star$ based on the mass -- excitation diagram \cite[MEx diagram,][]{Juneau+11}. 
In fact, assuming that our galaxy is star forming, the ratio between the optical [OIII] doublet and \hb\ implies a maximum stellar mass on the order of $10^{9.5} {\rm M}_\odot$.
This puts our target in the typical mass range of LAEs \cite[][]{Ouchi+20, Pucha+22} and lensed galaxies at similar redshifts \cite[e.g.][]{Mestric+22, Bouwens+22}.
Besides, this estimate agrees with the fact that low stellar mass galaxies tends to display larger SFR(\ha)/SFR(UV) ratios \cite[][]{Faisst+19, Atek+22}.

\begin{figure}
    \centering
    \includegraphics[width=\columnwidth, trim={0cm 0cm 0cm 1cm}]{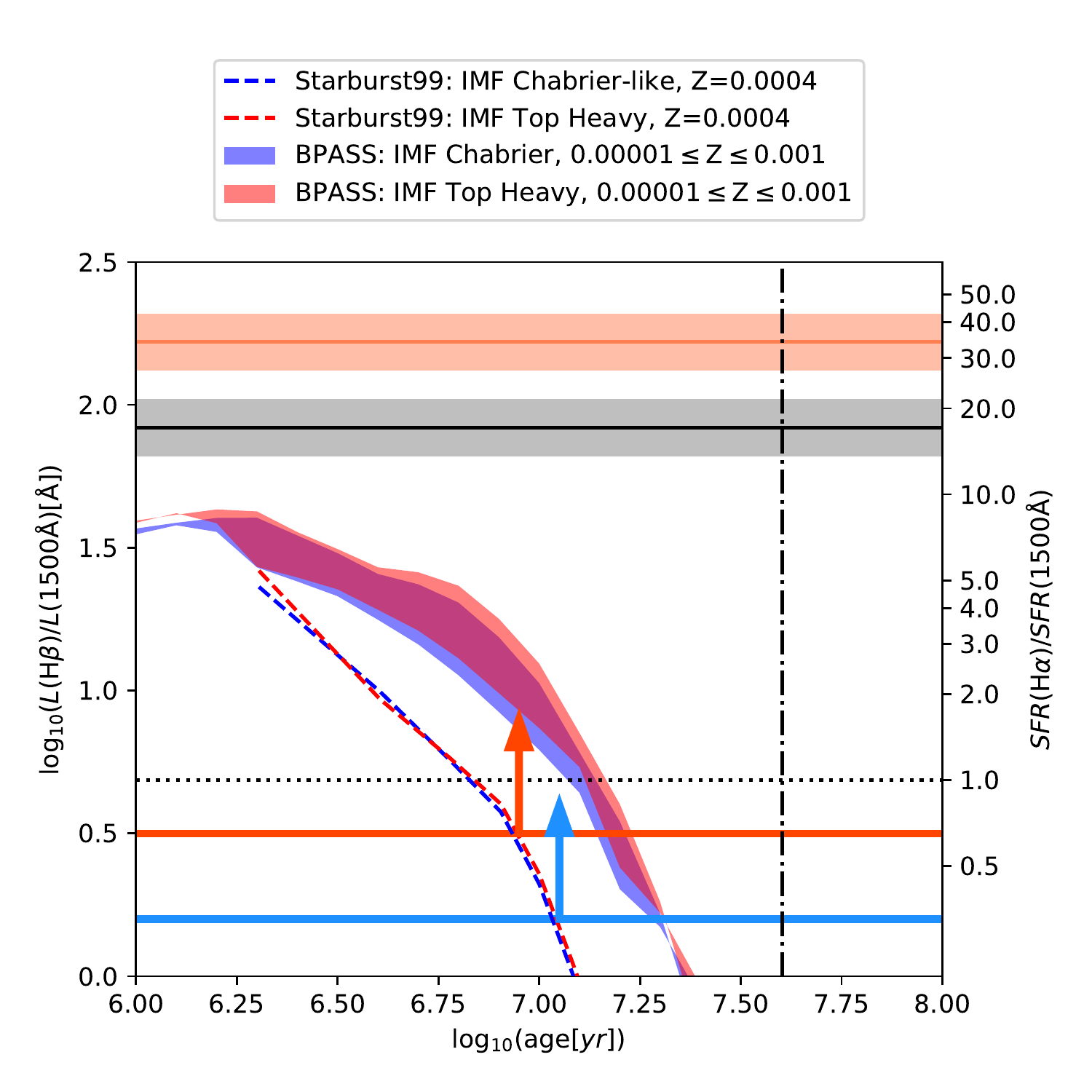}
    \caption{Age versus L(\hb)/L(UV) (left) and SFR(\ha)/SFR(1500\AA) (right) ratios diagram.
    With dashed lines we present the theoretical tracks from \textsc{Starburst99} for a Chabrier-like (blue) and top-heavy (red) IMFs. Following the same colour coding, the blue and red shaded areas depict the models from \textsc{bpass} v. 2.2.1. The horizontal black line shows the observed L(\hb)/L(UV) and SFR(\ha)/SFR(1500\AA) ratios for our target and the grey shaded region is representative of its associated error.
    Similarly, in orange, we show the range of values obtained if we correct the observed ratios for nebular continuum emission. The coloured arrows highlight the ratios lower-limits of the aforementioned estimates in the case of dust extinction correction (assuming the upper-limit value $E(B-V) = 0.75$) with the attenuation curve by \protect\cite{Calzetti+00}. The horizontal dotted black line is representative of SFR(\ha)/SFR(1500\AA$)=1$. Finally, with a vertical black dash-dotted line we report the upper-limit on the age of the galaxy youngest stellar population (40 Myr) if we consider the results from the analysis of the (C/O) abundance, see Section~\ref{subsec:te_ne}.}
    \label{fig:age_tracks}
\end{figure}

We estimate the light-weighted age of the stellar population through the ratio between the \hb\ and UV continuum luminosity. 
This ratio is predicted to decrease with increasing stellar age.
We derive theoretical tracks from both \textsc{Starburst99} and \textsc{bpass}.
For our target, the ratio between the observed luminosities is equal to $\log_{10}(L({\rm H}\beta)/L_\nu(1500\text{\AA})) = 1.92\pm0.10$, see Figure~\ref{fig:age_tracks}.
If we account for the contribution of the nebular continuum, the estimate raises to $2.22\pm0.10$. 
In both cases, the retrieved values are above the \textsc{Starburst99} and \textsc{bpass} theoretical predictions.
Only if we correct for reddening, our results become compatible with the tracks. 
Assuming the extreme value of $E(B-V) = 0.75$ (i.e. the upper-limit derived from the Balmer decrement \hg/\hb, see Section~\ref{subsec:beta_slope}), we obtain stellar ages $\lesssim 10-20$ Myr. 

\subsection{The \lya~emission}
\label{subsec:lya}
In the following section, we present a detailed description of the spectral and spatial properties of the \lya\ emission detected in our target.

\subsubsection{Spectral properties of the \lya}
\label{subsec:lya_fit}
The spectral shape of the \lya\ emission of our target is double peaked. We model each peak through the asymmetric Gaussian profile introduced by \cite{Shibuya+14b}. 
From the best-fit model, we obtain a \lya\ observed total flux of $(1.66\pm0.03)\cdot 10^{-17}$ erg/s/${\rm cm}^2$, corresponding to a total luminosity of $(2.14\pm0.02)\cdot 10^{42}$ erg/s. 
The flux ratio between the blue and red peak is of $\sim 14$ per cent.
Since we do not detect any stellar continuum underneath the \lya, we estimate a \lya\ EW$_0 \leq -108$\AA. 
This value is obtained by dividing the \lya\ flux by the median value of the 1-$\sigma$ error in the spectral region covered by the \lya\ line. 
The estimate is in agreement with the EW$_{0}$ of other \lya-emitters at similar redshifts and luminosities \cite[e.g.][]{Runnholm+20}. 
Despite the smooth spectral profile, we highlight that contributions from HeII$\lambda1215$ and the OV]$\lambda\lambda1214,1218$ doublet could partially affect the \lya\ estimated flux, luminosity and EW$_0$. 
Theoretical models show that the contribution of the HeII and OV] doublet varies as a function of several physical parameters: the power-law index of the ionising radiation, the nebular metallicity $Z_{\rm neb}$, the ionisation parameter $U$, and the presence and strength of HI absorptions \cite[][]{Humphrey+19}. 
According to these models and based on our target estimated nebular metallicity and ionisation parameter, the HeII and OV] lines could contribute from 0.4 (optically thick models) to 8 per cent (optically thin models) to the total \lya\ flux. 
For our object, in the case of an optically thin model, the contaminating flux would be mostly associated to the OV] doublet since the HeII$\lambda1215$ flux is negligible ($F({\rm HeII}\lambda1215) = (0.28-0.33)\times F ({\rm HeII}\lambda1640) \simeq 6.8\times 10^{-20}$ erg/s/cm$^2$, i.e. 0.4 per cent of the \lya\ flux).
However, since we do not have direct evidence of the OV] doublet in our spectrum and we cannot more robustly constrain its possible flux from the analysis of other lines \cite[e.g. from NV$\lambda1240$ or $\rm{[NeV]}\lambda1575$,][]{Humphrey+19}, in the following we do not consider corrections to the reported \lya\ flux, luminosity and EW$_0$.

The separation between the blue and red peak of the \lya\ emission is $\Delta {\rm v}_{\rm peak}=459\pm38$ km/s, with the red peak offset from the systemic velocity of $320\pm36$ km/s. 
Following the results by \cite{Kaikiichi+19}, the estimated peak separation suggests a Lyman-continuum escape fraction $f_{\rm esc}^{\rm LyC} \lesssim 15$ per cent. 
This is also confirmed by the trend recently reported by \cite{Schaerer+22} between the $f_{\rm esc}^{\rm LyC}$ and the C43 ratio (i.e. CIV$\lambda1550$/CIII$\lambda1906,1909$). 
According to their study, \cite{Schaerer+22} report that galaxies with ${\rm C43}<0.75$ typically have Lyman-continuum escape fraction below 10 per cent.
For our target, the observed C43 ratio is equal to $0.18\pm0.03$, thus reinforcing the conclusion that our target is a weak Lyman-continuum leaker.

For case B recombination and no dust extinction, the expected ratio \lya$/$\hb~$=31.9$\footnote{The reported value has been obtained with \textsc{PyNeb} assuming T$_e=2\times10^4$~K and $n_e = 10^4~{\rm cm}^{-3}$.}.  
We estimate \lya$/$\hb~$=2.4\pm0.5$, a factor of about 13 lower than the theoretical ratio, and that translates into a \lya\ escape fraction $f_{\rm esc}^{{\rm Ly}\alpha}$ (i.e. $L_{{\rm Ly}\alpha}^{\rm obs}/L_{{\rm Ly}\alpha}^{\rm int}$) of about 8 per cent. 
This result is in good agreement with the global \lya\ escape fraction typically observed at $z\sim 3-4$ \cite[e.g.][]{Hayes+11}. Besides, this estimate appears to be consistent with other objects with similar EW$_0$ \cite[e.g.][]{Erb+16} and with the $f_{\rm esc}^{\rm Ly\alpha} - {\rm EW}_0$ correlation after having properly taken into account the impact of the target's ionising efficiency $\xi_{\rm ion}$ \cite[$\xi_{\rm ion} = 1.3\times 10^{25}\times {\rm SFR(H\alpha)/SFR(UV)}$~Hz/erg; see][and their Equation 6, for further details]{Sobral-Mathee+19}. 
Interestingly, if we hypothesise that the tension between the theoretical and the observed \lya$/$\hb\ ratios is only due to dust extinction, we would estimate a colour excess $E(B-V)_{{\rm Ly}\alpha} = 0.38\pm 0.03$ (adopting the attenuation curve by \citealt{Calzetti+00}) quite in agreement with the nebular colour excess $E(B-V)_{\rm neb}$ that we estimated from the \hg$/$\hb\ ratio ($0.34^{+0.41}_{-0.34}$, see Section~\ref{subsec:beta_slope}).

We apply radiative transfer modelling to the \lya\ emission to infer the ISM metal and dust content \cite[e.g.][]{Charlot+93}, HI and HII regions relative geometries, the kinematics of the neutral gas \cite[e.g.][]{Verhamme+06,Laursen+07} directly from the profile of the \lya\ line.
We apply an updated version of the pipeline introduced by \cite{Gronke+15}, implementing 12960 radiative transfer models derived by means of the code \texttt{tlac} \cite[][]{Gronke+14}. The radiative transfer models we adopt are ‘shell-models’ \cite[e.g.][]{Ahn+02,Verhamme+06}, i.e. they consist of a single point-like source emitting \lya\ and continuum radiation and surrounded by a shell of neutral hydrogen, and dust. Because of their simple geometry, only four parameters are needed to fully describe these models: the shells neutral hydrogen column density $N_{\rm HI}$, their expansion velocity ${\rm v}_{\rm exp}$ (assuming positive values for outflows, negative in the case of inflows), an (effective) temperature T that includes the effects of small-scale turbulence, and the dust optical depth $\tau_d$ to parametrise the dust content. Finally, the emitted \lya\ is assumed to have an intrinsic Gaussian emission characterised by an intrinsic \lya\ equivalent width ${\rm EW_{int}}$, and its width $\sigma_{\rm int}$.
We carry out the fitting in wavelength space with a Gaussian prior on
the redshift $z$ and after degrading the synthetic spectrum to the MUSE spectral resolution at the observed \lya\ wavelength
(derived from equation 8 in \citealt{Bacon+17}). 
According to the best-fit model, for the shell we obtain $\log_{10}(N_{\rm HI}{\rm [cm^{-2}]}) = 16.8\pm0.2$, ${\rm v}_{\rm exp} = 99\pm5$ km/s, $\log_{10}({\rm T [K]})=5.3\pm0.2$ and $\tau_d = 0.6\pm0.2$, while the \lya\ intrinsic emission has ${\rm EW_{int}} = - 415\pm12$ \AA\ and $\sigma_{\rm int} = 210\pm4$km/s.
As already pointed out in recent literature \cite[e.g.][]{Orlitova+18}, the intrinsic width of \lya\ recovered by the radiative transfer models results to be significantly larger than the one of the Balmer lines (for our target a factor $\sim 3.4$ larger if compared to our estimate of the \hb\ line, in line with the results from \citealt{Orlitova+18}). 
According to \cite{Li+22}, however, this discrepancy is most likely due to additional radiative transfer effects and, in particular, the possibility that the \lya\ propagates within a clumpy gas distribution with velocity dispersions $\gtrsim 100$ km/s.

Since 'shell-models' are an oversemplification of the complex structure and kinematics of \lya-emitters and their surroundings, it is currently debated how much of the radiative transfer is indeed captured by the models and the real meaning and reliability of the involved parameters, especially T and $\tau_d$. At the same time, it has been shown that the outflow velocity ${\rm v}_{\rm exp}$ and column density of the ‘shell-model’ $N_{\rm HI}$ correlate well with the ones of a more realistic multi-phase medium \cite[][]{Gronke+16}, thus making the estimate of these two parameters particularly robust.
We report the results obtained from the analysis of the \lya\ spectral properties in Table~\ref{tab:lya_params}.

Finally, we investigate the presence of spatial variations and kinematics patterns in the \lya\ spectral shape \cite[e.g.][]{Patricio+16, Smit+17, Erb+18, Claeyssens+19, Leclercq+20}.
To this aim, we extract the 1st moment map of the \lya\ spectral distribution as well as the 0th moment maps created by collapsing the MUSE datacube in velocity bins around the \lya\ line. 
From both methods, we find no coherent kinematic pattern.
To increase the SNR of the profiles, we decide to compare the integrated \lya\ spectrum extracted within a circular aperture ($0.4''$ radius) centred at the peak position of the \lya\ emission with the \lya\ profile observed in the outskirts (circular annulus $0.4''-0.8''$).
Also in this case, we do not find any difference in the \lya\ profile shape of the two regions and in contrast with results from recent studies of other lensed LAEs \cite[e.g.][]{Claeyssens+19, Chen+21}.

\begin{table}
    \centering
    \begin{tabular}{lr}
    \hline
    \hline
    Parameter  & Value \\
    \hline
    $F({\rm Ly}\alpha)$  [erg/s/cm$^2$]       & $(1.66\pm0.03)\times10^{-17}$\\
    $L({\rm Ly}\alpha)$  [erg/s]              & $(2.14\pm0.02)\times10^{42}$\\
    $F_{\rm blue,peak}/F_{\rm red,peak}$      & $\sim 14 \%$\\
    EW$_0$ [\AA]                              & $\leq -108$\\. 
    $\Delta {\rm v}_{\rm tot}$ [km/s]               & $459\pm38$\\
    $\Delta {\rm v}_{\rm red,peak}$ [km/s]          & $320\pm36$\\
    $f^{{\rm Ly}\alpha}_{\rm esc}$            & $0.08\pm0.02$\\
    $f^{{\rm LyC}}_{\rm esc}$                 & $<15$\%\\
    $\log_{10}(N_{\rm HI} {\rm[cm^{-2}]})$    & $16.8\pm0.2$\\
    ${\rm v}_{\rm exp}$ [km/s]                      & $99\pm 5$\\
    $\log_{10}(T~[\text{K}])$                 & $5.3\pm0.2$\\
    $\tau_d$                                  & $0.6\pm0.2$\\
    ${\rm EW_{int}}$ [\AA]                    & $-415\pm12$\\
    $\sigma_{\rm int}$ [km/s]                 & $210\pm4$\\
    \hline
    \end{tabular}
    \caption{Table of the results from the line fitting procedure and \texttt{tlac} radiative transfer modelling of the target \lya\ emission, see Section~\ref{subsec:lya_fit}. The line flux $F({\rm Ly}\alpha)$ and luminosity $L({\rm Ly}\alpha)$ are corrected for magnification but not for dust extinction.}
    \label{tab:lya_params}
\end{table}

\subsubsection{Spatial extent of the \lya}
\label{subsec:lya_spatial}

\begin{figure*}
    \centering
    \includegraphics[width=\columnwidth]{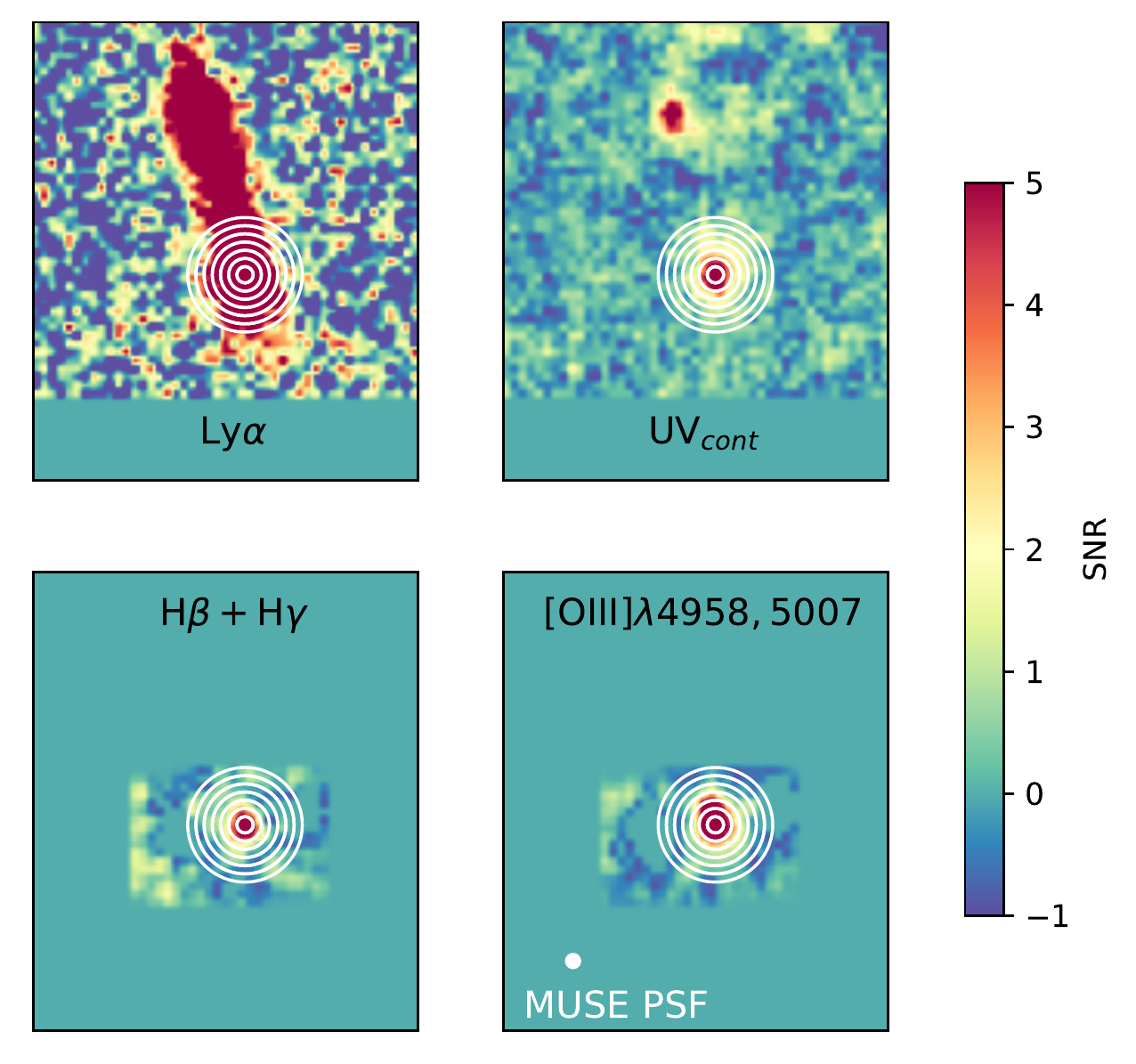}
    \includegraphics[width=\columnwidth]{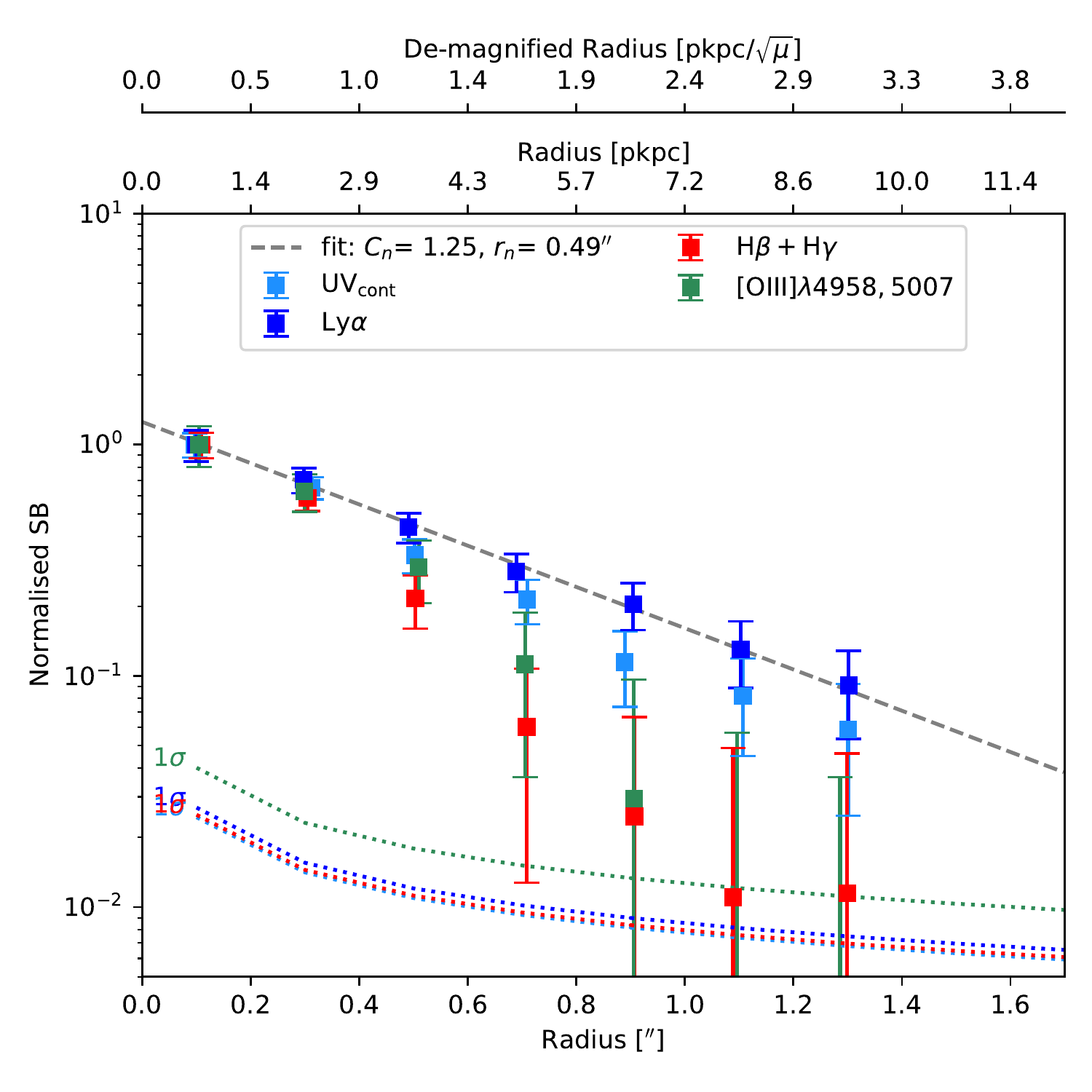}
    \caption{\textbf{Left panels}: SNR maps of the \lya, UV continuum, Balmer lines and optical [OIII] doublet of our target. The concentric white squares show the areas within which the SB profile of the different tracers (bottom panels) has been extracted. In the left bottom corner of the left-hand side panel, we also report with a white circle the size of the MUSE PSF. \textbf{Right panel}: radial surface brightness profiles of the \lya, UV continuum and strongest optical emission lines of our target. The measurements (coloured squares) are normalised to their peak value. The grey dashed line shows the best-fit exponential profile to the \lya\ data-points (blue squares), while the coloured dotted lines shows the $1\sigma$ limit for each tracer.}
    \label{fig:sb_profiles}
\end{figure*}

Following the procedure described in Section~\ref{subsec:nb-images_spectra}, we extract the pseudo-NB images of the strongest emission lines of our target, see top panels in Figure~\ref{fig:sb_profiles}.
We also create a pseudo-NB image of the galaxy UV continuum emission by collapsing the MUSE datacube in the wavelength windows within which we measured the UV $\beta$-slope, see Table~\ref{tab:spectral_windows} Section~\ref{subsec:beta_slope}, and stacking them together.
Similarly, to increase the SNR of the emission line images we stack the intensity maps of the doublets of the optical [OIII]. 
We also stack together the maps of the two Balmer transitions available, i.e. \hb\ and \hg.
Because of the different spatial resolution of the MUSE and SINFONI data, we degrade the resolution of the SINFONI pseudo-NB images to match the one of MUSE. 
To do so, we first convolve the SINFONI images with a Gaussian kernel to match the MUSE PSF, and then we resample them to the MUSE standard spatial sampling, i.e. $0.2''\times0.2''$. 

Several statistical studies have shown the existence of an offset between the centroid of the \lya\ and UV continuum emission $\delta_{\rm Ly\alpha}$ \cite[e.g.][]{Bunker+00, Fynbo+01, Shibuya+14a, Hoag+19, Ribeiro+20, Claeyssens+22}. 
Theoretical studies based on the 3D modelling of \lya\ radiative transfer in disk galaxies \cite[e.g.][]{Laursen+07, Verhamme+12, Zheng+14} ascribe the \lya\- UV offset to the easier escape and propagation of \lya\ photons perpendicularly to the galaxy disk. 
In this case, the presence of an offset could be explained as a consequence of the viewing angle under which the observer sees the target. 
Interestingly, \cite{Claeyssens+22} report how the presence of an offset between the galaxy UV emission and the \lya\ could also be linked to the mechanism originating the \lya\ emission. In the case of small offsets (whenever the \lya\ centroid is within the effective radius of the UV emission), \cite{Claeyssens+22} suggest that the offset likely originates from substructures detectable in the UV, such as an off-center star-forming clump. In the case of larger offsets, other mechanisms could be invoked: gas inflows, scattering effects of the \lya\ photons in the CGM, extinction, fluorescence, emission from faint satellites. To investigate this, we resort to the \textsc{galfit} software \cite[][]{Peng+10}. 
We model the multiple images M2 and M3 of our target simultaneously in both the \lya\ and UV continuum map. 
For each multiple image, we use a single S\'ersic profile and obtain smooth residual maps (normalised residuals $< 20$\%).
From the \textsc{galfit} best-fit, we estimate an offset between the centroids of the \lya\ and UV continuum $\delta_{\rm Ly\alpha}=0.18\pm0.02$ arcsec that corresponds to $1.30\pm0.17$ pkpc. If we correct for magnification\footnote{To correct for magnification, we divide the estimated $\delta_{\rm Ly\alpha}$ value for the square root of the magnification factor $\mu = 9\pm 2$ (see Table 2 in \citealt{Livermore+15}).}, the intrinsic offset results in $0.43\pm0.07$ pkpc. 
This estimate is in very good agreement with the recent results presented by \cite{Claeyssens+22} who found a median value for the separation of the \lya\ and UV continuum centroids of $\delta_{\rm Ly\alpha}=0.58\pm0.14$ pkpc in the analysis of 603 lensed \lya-emitters in the redshift interval $2.9<z<6.7$. 
In addition, our estimate is also in line with the findings by \cite{Hoag+19} and \cite{Ribeiro+20} for \lya-emitters at similar redshifts. 
Besides, the low $\delta_{\rm Ly\alpha}$ value is in trend with the observed \lya\ EW$_0$ -- $\delta_{\rm Ly\alpha}$\ anti-correlation \cite[e.g.][]{Shibuya+14b, Hoag+19}. 
Hence, following \cite{Claeyssens+22}, the observed offset in our target could be the consequence of an off-centre star-forming clump harboured in our target.
 
Additionally, we investigate the surface brightness (SB) profiles of UV lines and continuum. From each pseudo-NB image, we extract the radial SB profile within concentric apertures centred at the peak of the UV continuum emission and with increasing radii (in steps of $0.2''$) from $0.2''$ out to $1.4''$ (Figure~\ref{fig:sb_profiles}).
The UV and optical lines and the UV continuum appear to be all spatially resolved, having a larger spatial extent than the MUSE PSF. 
In particular, the \lya\ and UV continuum are the most spatially extended.
By fitting the \lya\ SB with an exponential function, i.e. SB$(r)= C_n \exp{(-r/r_n)}$, we obtain a scale length $r_n$ for the \lya\ nebula of $r_n({\rm Ly}\alpha) = (0.49\pm 0.01)''$.
If we convert into physical kpc and correct for magnification, we infer $r_n({\rm Ly}\alpha)\simeq1.2$ pkpc (not corrected for PSF). 
This value is in good-agreement with the typical scale length estimates for \lya\ halos and with the $r_n$ -- $L$(\lya) correlation \cite[e.g. figure 13 in][and references therein]{Ouchi+20}. 
If we apply the exponential model to the UV continuum and Balmer lines we derive $r_n({\rm UV}) = (0.38\pm 0.02)'' \simeq 1$pkpc and $r_n({\rm H}) = (0.25\pm 0.02)'' \simeq 0.65$pkpc.
According to the $r_n$ values and the trend of the SB profiles presented in Figure~\ref{fig:sb_profiles}, the \lya\ and UV continuum extend more than the Balmer lines.

\section{Summary and Discussion}
\label{sec:discussion}
The multi-wavelength analysis of our target highlights that A2895b is a lensed Lyman-$\alpha$ emitter at $z\simeq 3.721$ with a compact UV morphology ($r_n \simeq 1.2$~pkpc) and a \lya\ luminosity of $\simeq 2\times 10^{42}$ erg/s.
In particular, A2895b is a star-forming galaxy with a SFR estimated from the conversion of the \hb\ luminosity of $10.7\pm 2.3$~\msunyr. 
The MEx diagram \cite[][]{Juneau+11} suggests that our target has a stellar mass smaller than $10^{9.5} M_\odot$. 
If we consider the upper-limit on the stellar mass and the above SFR, A2895b populates a region of the ${M_\star}$ -- SFR plane above the main-sequence of star-forming galaxies at $2.8<z<4$ \cite[][]{Rinaldi+21}, with a lower-limit on its specific SFR of about $3.5 \times 10^{-9}~{\rm yr}^{-1}$.  
This suggest that our target is likely to be a starbursting system.

From the analysis of the nebular UV line ratios, their shape and width, and the absence of an X-ray counterpart (see Section~\ref{subsec:agn_or_sfr}), we exclude the presence of an AGN.
In this regard, the diagnostic diagrams based on UV line ratios clearly show a net separation between the typical values observed in high-$z$ radio galaxies \cite[e.g.][]{Vernet+01, Humphrey+08} and type-II quasars \cite[][]{Silva+20}, as well as from theoretical predictions of AGNs narrow-line regions \cite[][]{Feltre+16}.
This result is also in line with the fact that LAEs hosting AGNs are typically characterised by a \lya\ luminosity ${\rm L(Ly\alpha)} \gtrsim 2.5 \times 10^{43}$~erg/s, i.e. one order of magnitude above our estimate \cite[e.g.][]{Konno+16}. 

Based on these findings, in-situ star-formation results to be a good candidate for explaining the \lya\ emission of our target, whereas we disfavour AGN activity.
We also exclude scenarios of both shock heating due to outflows and gravitational cooling.
In fact, these processes not only are typically expected to contribute to the \lya\ emission at $\gtrsim 20$~pkpc from the galaxy centre \cite[e.g.][and references therein]{Mas-Ribas+17}, but to significantly increase the number of collisional excitations of the hydrogen atoms, thus bringing to an increase in the measured \lya/\hb\ ratio with respect to what predicted by case B recombination \cite[\textit{super case B objects}, e.g.][]{Oti-Floranes+12, Nakajima+13}.
Specifically, in the case of a significant contribution to the \lya\ emission from inflowing gas accreted onto the central galaxy \cite[i.e. gravitational cooling, e.g.][]{Haiman+00, Keres+05, Dekel+06, Shull+09}, the \lya/\hb\ ratio is expected to exceed the case B value of more than a factor of 10 \cite[e.g.][]{Dijkstra+14}. 
Nonetheless, for our target we estimate a \lya/\hb\ ratio 13 times smaller than the case B value, i.e. 1-2 orders of magnitudes below the predictions for shock heating and gravitational cooling (see Section~\ref{subsec:lya_fit}).
Besides, for the gravitational cooling hypothesis, signatures of \lya\ cooling radiation should be found by the analysis of the \lya\ EW.
In fact, in this case the \lya\ intrinsic EW is expected to exceed the maximum predictions for `regular' star-formation activity \cite[i.e. $300-400$~\AA\ in modulus, e.g.][]{Kashikawa+12}.
Also in this regard, we do not need to resort to gravitational cooling to explain our measurements (${\rm EW}_{\rm int} = -415\pm 12$~\AA).
We highlight, however, that the details of the \lya\ emission via gravitational cooling are still debated and remain uncertain \cite[][]{Yang+06, Dijkstra+09, Faucher-Giguere+10, Cantalupo+12, Lake+15}.
Ultimately, we discard the hypothesis of fluorescence due to ex-situ star-formation activity (e.g. from satellite sources), or background sources (QSOs), since we do not find evidence of spectral features of these additional systems in our UV and optical spectra.

The analysis of the spatial extent of the UV continuum, \lya\ and Balmer lines (see Section~\ref{subsec:lya_spatial}) arises some interesting points.
For our target, the UV continuum and \lya\ emissions are more extended than the optical hydrogen transitions \hb$+$\hg.
While a more extended \lya\ SB profile with respect to the Balmer lines could be the consequence of resonant scattering, the fact that the UV continuum is more extended than the Balmer lines is more puzzling and seems to further provide evidence against fluorescence.
In fact, while nebular UV continuum emission follows recombination processes in the ISM, Balmer lines arise from both recombination and fluorescence. Hence, we would expect the nebular UV continuum emission to be more compact than the Balmer lines.
A possible way to explain this finding is to consider that the extended UV emission observed originates from an unobscured stellar population with age $20 - 100$~Myr. 
The existence of such stellar population seems to be supported by the very blue $\beta$-slope of our target ($\beta=-2.6\pm0.5$, see Section~\ref{subsec:beta_slope}).
If so, A2895b would be hosting (at least) two stellar populations: a newly born population of stars ($\leq 10$ Myr) traced by the Balmer lines, and a more extended, `older' and unobscured stellar population.
In this case, the extended \lya\ emission could arise from resonant scattering of the \lya\ photons produced in the star-forming regions.
Interestingly, the presence of two populations could also alleviate the discrepancy of the face value between the dust colour extinction estimated from the UV continuum and Balmer decrement \cite[e.g. \textit{dust selective extinction},][]{Calzetti+94}, along with the partial remaining tension between the ratio L(\hb)/L(UV) and the theoretical stellar models from \textsc{bpass} and \textsc{Starburst99} calculated in the case of top-heavy IMF and sub-solar metallicities, see Section~\ref{subsec:mass_age}.  
To verify this hypothesis, deeper observations targeting the galaxy rest-frame optical and IR emission are needed.
Such follow up could also explain the small \lya\ offset ($\delta_{\rm Ly\alpha} \simeq 0.6$~pkpc) detected with respect to the UV continuum and Balmer lines.
In fact, if $\delta_{\rm Ly\alpha}$ is a consequence of an off-centre star-forming clump \cite[e.g.][]{Claeyssens+22}, we would expect the \lya\ peak to be coincident with the peak of the Balmer lines.
However, this discrepancy could be a mere effect of how the astrometry of the SINFONI dataset has been registered to MUSE.
As explained in Section~\ref{subsec:sinfoni_reduction}, we registered the SINFONI astrometry to the MUSE one by minimising the spatial offset between the centroid of the optical [OIII] emission and the target UV continuum. 
By doing so, we also overlapped the centre of the optical and UV [OIII] emissions.
We followed this procedure since no other target falls within the SINFONI FoV and the optical continuum of our source is undetected. 
In this sense, rest-frame optical observations with a more robust astrometry could shed light on this aspect.

\section{Conclusions}
\label{sec:conclusions}
In this paper, we presented a detailed study of the integrated UV and optical properties of A2895b, a lensed \lya-emitter at $z\simeq3.721$ in the background of the A2895 galaxy cluster.
The analysis was based on the AO-assisted integral field spectroscopy of MUSE and SINFONI.
From our study, we inferred that:
\begin{enumerate}
    \item our target has a steep blue UV continuum ($\beta=-2.6\pm0.5$). Such blue continuum is possibly the sign of a young, unobscured stellar population.
    
    \item the analysis of the shape of spectral lines along with empirical diagnostic diagrams based on UV line ratios (C4C3-C34, C3-O3, He2-O3C3) suggests that our target is a star-forming galaxy with a current star-formation rate ${\rm SFR} = 10.7\pm2.3$~\msunyr\ \cite[from the \hb\ line luminosity,][]{Kennicutt+98}. 
    We exclude the presence of an AGN harboured in the galaxy centre.
    
    \item the galaxy has a sub-solar nebular metallicity ${\rm Z} = 0.05 \pm 0.02 {\rm Z}_\odot$ \cite[He2-O3C3 diagram][]{Byler+20} and (C/O) abundance ($\simeq 0.23 {\rm (C/O)}_\odot$). While the measured metallicity suggests a short star formation history, the low (C/O) value can be explained by both an over-abundance of massive O-type stars, hence, a top-heavy IMF, and a very recent burst of star-formation ($<< 40$ Myr).

    \item we find a significant discrepancy between the luminosity of the UV continuum at 1500 \AA\ and the \hb\ line. Stellar models (e.g. \textsc{bpass}, \textsc{Starburst99}) seem to explain this tension only in the case of a young stellar population ($< 10$ Myr) with low-metallicity (${\rm Z}\simeq 10^{-4}$) and that formed following a top-heavy IMF.
    To some extent, the introduction of dust extinction tends to alleviate the tension even further.
    \item the \lya\ emission of our target is double-peaked and has a total luminosity (not corrected for extinction) of $\simeq 2\times10^{42}$~erg/s.  
    The low value of the observed \lya/\hb/ ratio (\lya/\hb~$= 2.4\pm 0.5$) with respect to case B recombination, tends to exclude the hypothesis that the galaxy \lya\ emission originates from outflows (shock heating) and gravitational cooling. 
    This is also supported by the \lya\ intrinsic equivalent width \cite[obtained by modelling the \lya\ emission via radiative transfer `shell-models', e.g.][]{Gronke+14} that is in line with the typical values in the case of star-formation (i.e. $300 - 400$~\AA\ in modulus).
    \item The \lya\ is offset ($\delta_{\rm Ly\alpha} \simeq 0.6$~pkpc) from the UV continuum and Balmer lines.  The spatial extent of the \lya\ emission is comparable to the UV continuum while it is more extended than the Balmer emission (a factor about 2). The different extent of the emissions seems to suggest that while the UV emission traces a more extended region inhabited by an older and less extincted stellar population (10-100 Myr), the Balmer lines arise from regions of on-going star-formation ($\lesssim 10$~Myr). This could happen if phenomenon of fluorescence of \hb\ and \hg\ can be neglected, i.e. in the case of a low escaping fraction of ionising radiation from the production regions of the Balmer lines. In this scenario, the \lya\ extended emission would be just the consequence of resonant scattering from its region of production, an HII region. 
\end{enumerate}

The work presented in this paper shows the power of a multi-wavelength analysis in the characterisation of galaxy properties.
In particular, how the combination of rest-frame UV and optical information can help in tackling long-standing issues such as the origin of the \lya\ emission within galaxies.
So far, technical limitations (e.g. coarse spatial resolution, depth, limited field-of-view) have significantly limited such studies to small samples of objects \cite[e.g.][]{Nakajima+13, Erb+16, Trainor+19, Runnholm+20a, Weiss+21, Matthee+21, Reddy+22, Pucha+22}.
In this regard, the James Webb Space Telescope (\textit{JWST}), and in particular the on-board Near Infrared Spectrograph \cite[NIRSpec,][]{Jakobsen+22}, will be able to extend these studies to statistical samples of galaxies at intermediate redshifts, thus possibly providing us some definitive answers.


\section*{Data availability}
The data underlying this article will be shared on reasonable request to the corresponding author.

\section*{Acknowledgements}
The authors thank the anonymous referee for the useful comments and suggestions.
This work is based on observations collected at the European Southern Observatory under ESO programmes 087.B-0875(A), 60.A-9195(A) and 0102.B-0741(A).
This work made use of v2.2.1 of the Binary Population and Spectral Synthesis (\textsc{bpass}) models as described in Eldridge, Stanway et al. (2017) and Stanway, Eldridge et al. (2018).




\bibliographystyle{mnras}
\bibliography{biblio} 




\appendix
\section{Detection of additional \lya-emitters}
\label{sec:additional_laes}
During the analysis of the MUSE dataset in our hands, we detected five LAEs in addition to our target and the already known system at $z\sim3.4$ \cite[e.g.][]{Livermore+15, Iani+21}.
Among the newly discovered sources, we found two multiply-imaged systems (at $z\simeq 4.65$ and $z\simeq4.92$), and three single-imaged LAEs (at $z\simeq 4.57$ and $z\simeq4.92$).
According to the shape of their \lya\ emission and their spatial position, we exclude the hypothesis that the single-imaged LAEs at $z\simeq4.92$ are additional images of the multiply-imaged system at the same redshift.
In table~\ref{tab:additional_laes} we provide the spatial coordinates of these targets and a rough estimate of their redshift. 
For the nomenclature of these targets, we follow the naming adopted by \cite{Livermore+15}. 
We present MUSE cutouts for the \lya\ and UV continuum emission of each new target and related multiple images, as well as a zoomed-in image of their UV spectrum around the \lya\ line.  

\begin{table*}
    \centering
    \begin{tabular}{lccccc}
    \hline
    \hline
    LAE & $z$ & M1 & M2 & M3 & \hst-detected\\
    &  & $\alpha_{\rm J2000.0}$\ \ \ \ \ \ \ \ $\delta_{\rm J2000.0}$ & $\alpha_{\rm J2000.0}$\ \ \ \ \ \ \ \ $\delta_{\rm J2000.0}$ & $\alpha_{\rm J2000.0}$\ \ \ \ \ \ \ \ $\delta_{\rm J2000.0}$ & \\
    \hline
    A2895a & 3.395 & 01:18:11.190  -26:58:04.40 & 01:18:10.890  -26:58:07.50 & 01:18:10.570  -26:58:20.50 & y \\
    A2895b & 3.721 & 01:18:11.127  -26:57:59.36 & 01:18:10.543  -26:58:10.56 & 01:18:10.439  -26:58:14.36 & y \\
    A2895c & 4.57 & 01:18:09.151  -26:57:47.82 & - & - & n \\
    A2895d & 4.65 & 01:18:11.127  -26:57:59.56 & 01:18:10.304  -26:58:10.96 & - & n \\
    A2895e & 4.92 & 01:18:10.887  -26:57:58.56 & 01:18:10.783  -26:57:58.96 & 01:18:10.319  -26:58:08.36 & n \\
    A2895f & 4.92 & 01:18:09.465  -26:57:51.82 & - & - & n \\
    A2895g & 4.92 & 01:18:12.532  -26:58:38.82 & - & - & n \\
    \hline
    \end{tabular}
    \caption{Approximate redshift and spatial coordinates of the LAEs serendipitously detected in the MUSE observations of the A2895 galaxy cluster. We also report the possible presence of a possible counterpart in the available \hst\ ACS/WFC F606W image.}
    \label{tab:additional_laes}
\end{table*}

\begin{figure*}
\includegraphics[height=.32\textheight]{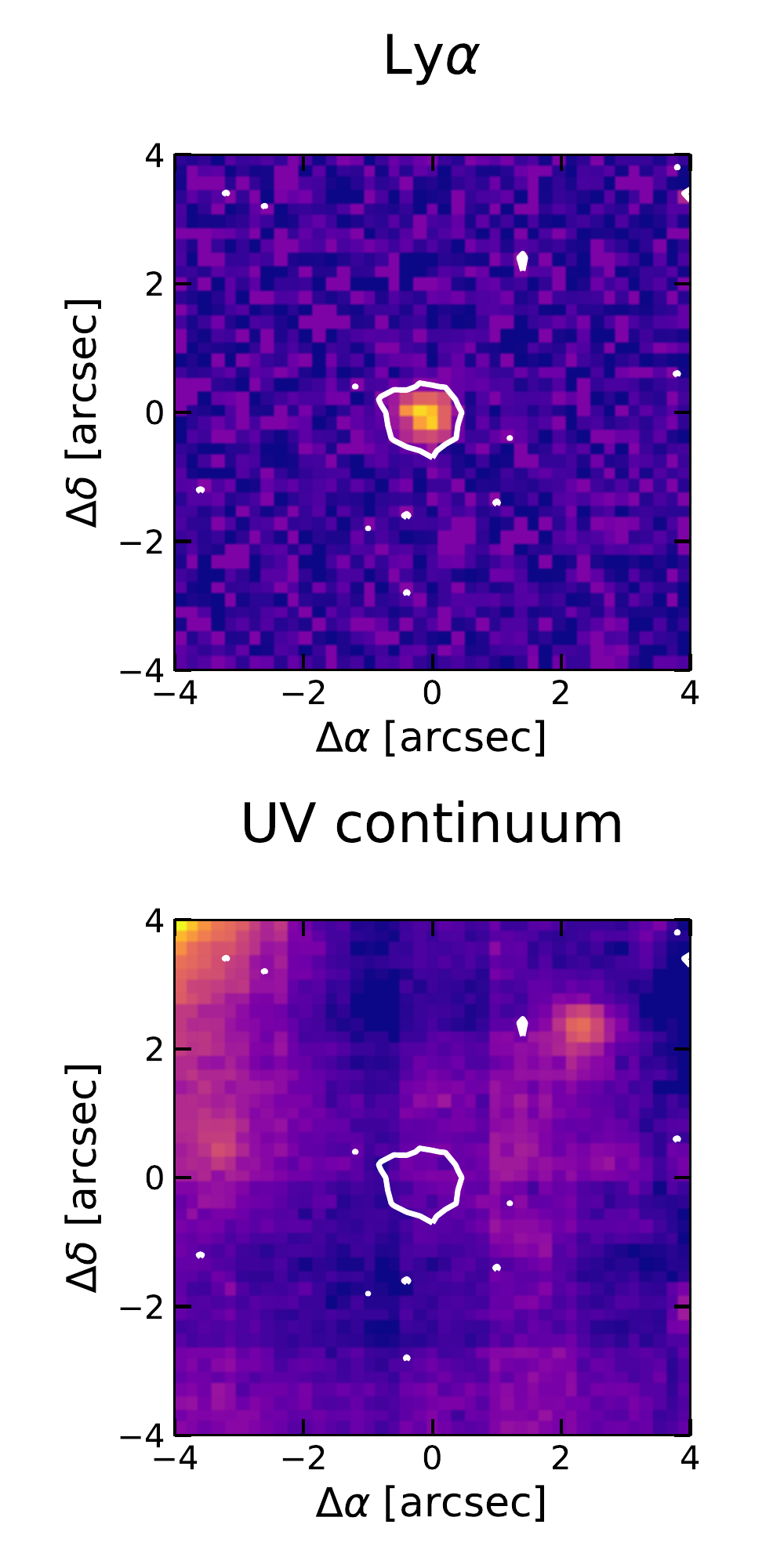}
\includegraphics[height=.32\textheight]{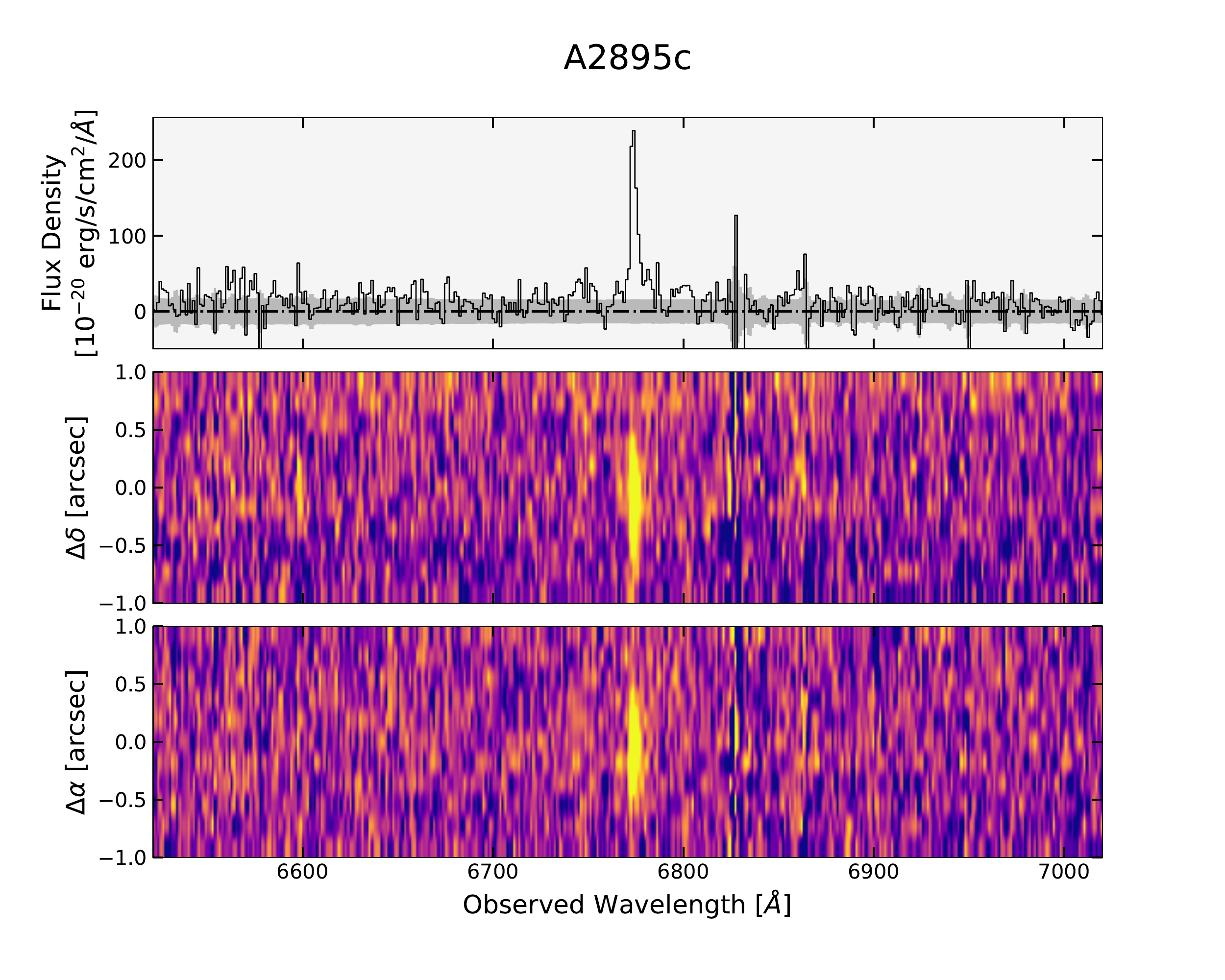}
\end{figure*}

\bigskip

\begin{figure*}
\includegraphics[height=.32\textheight]{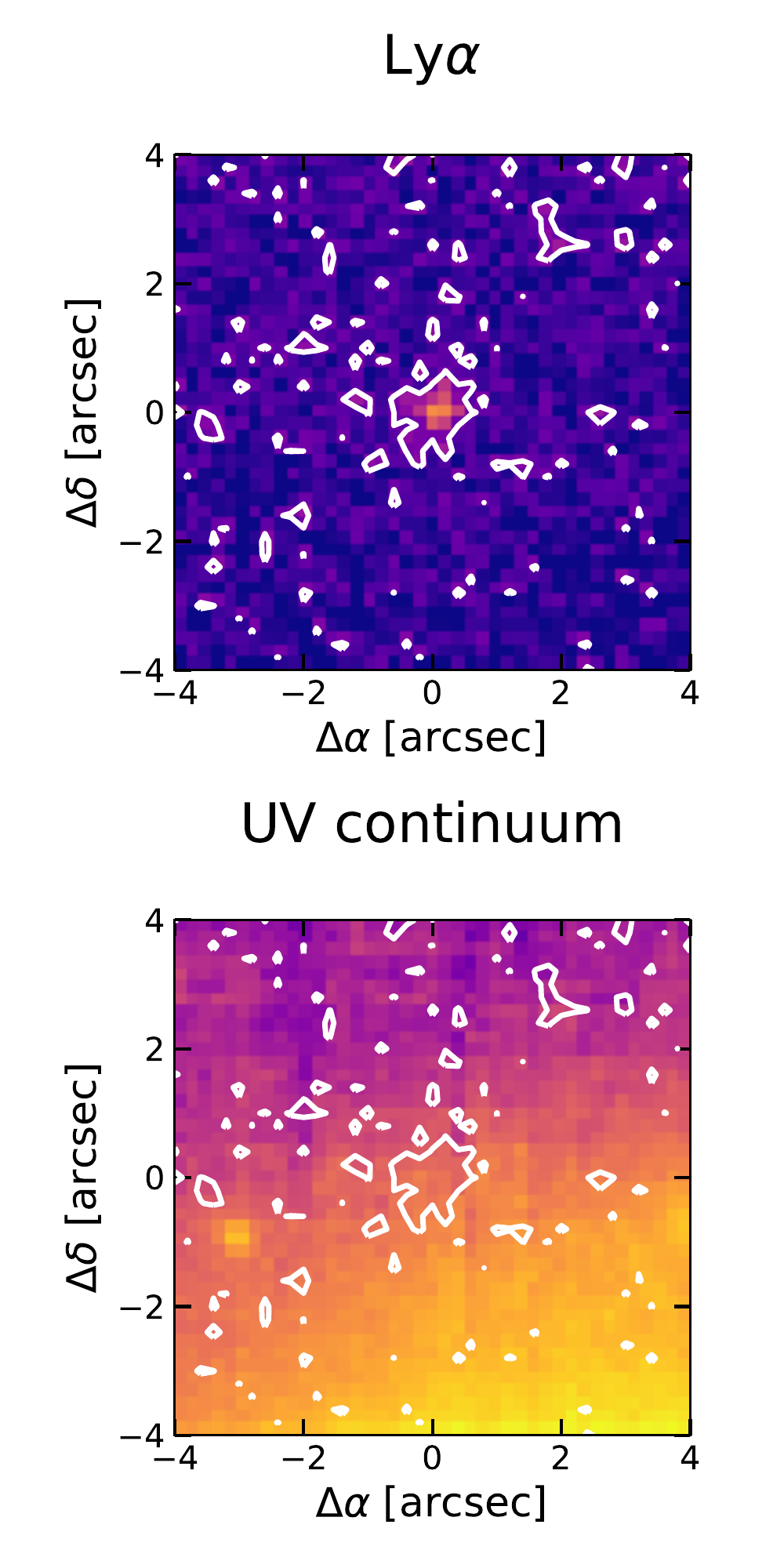}
\includegraphics[height=.32\textheight]{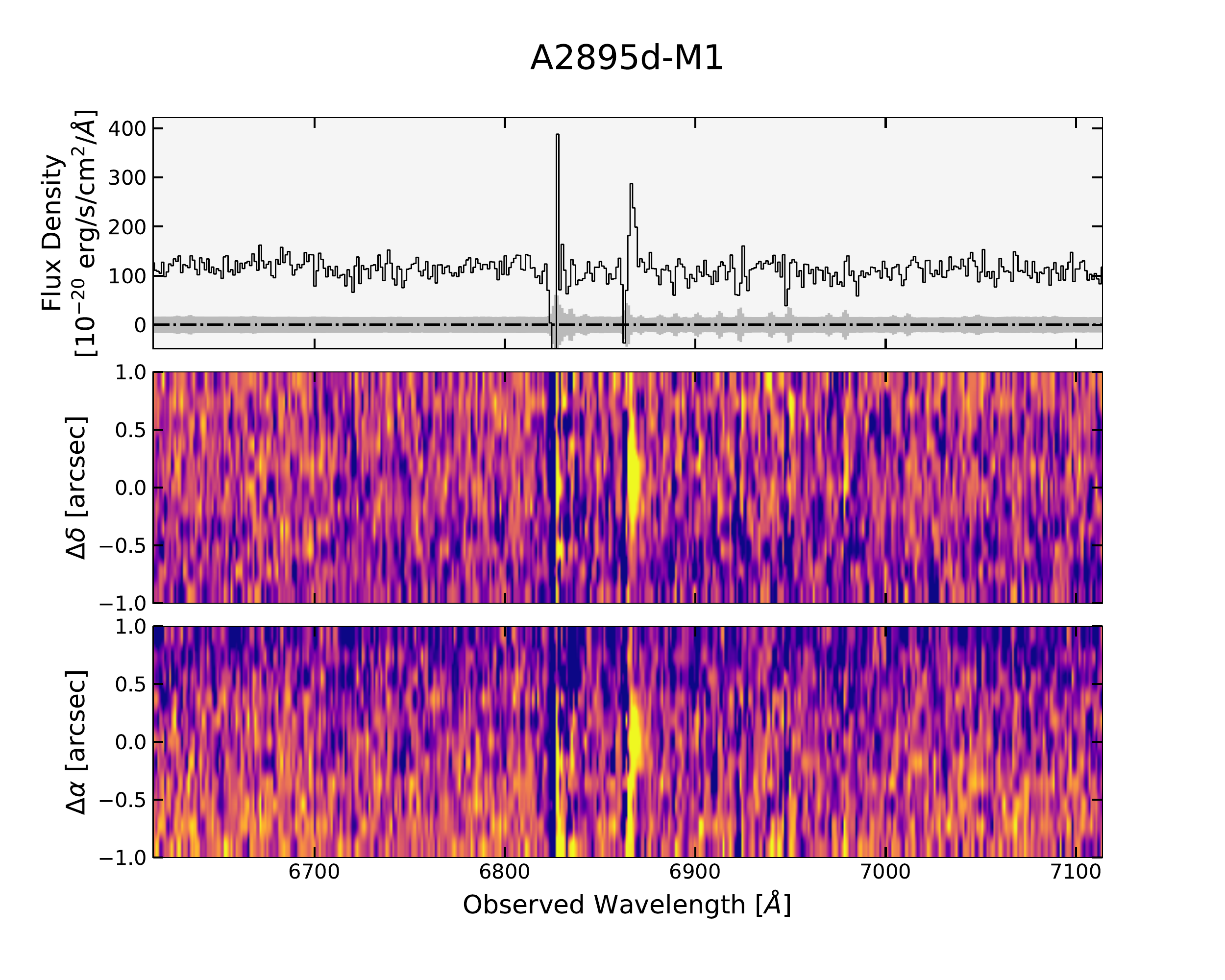}\\
\bigskip
\includegraphics[height=.32\textheight]{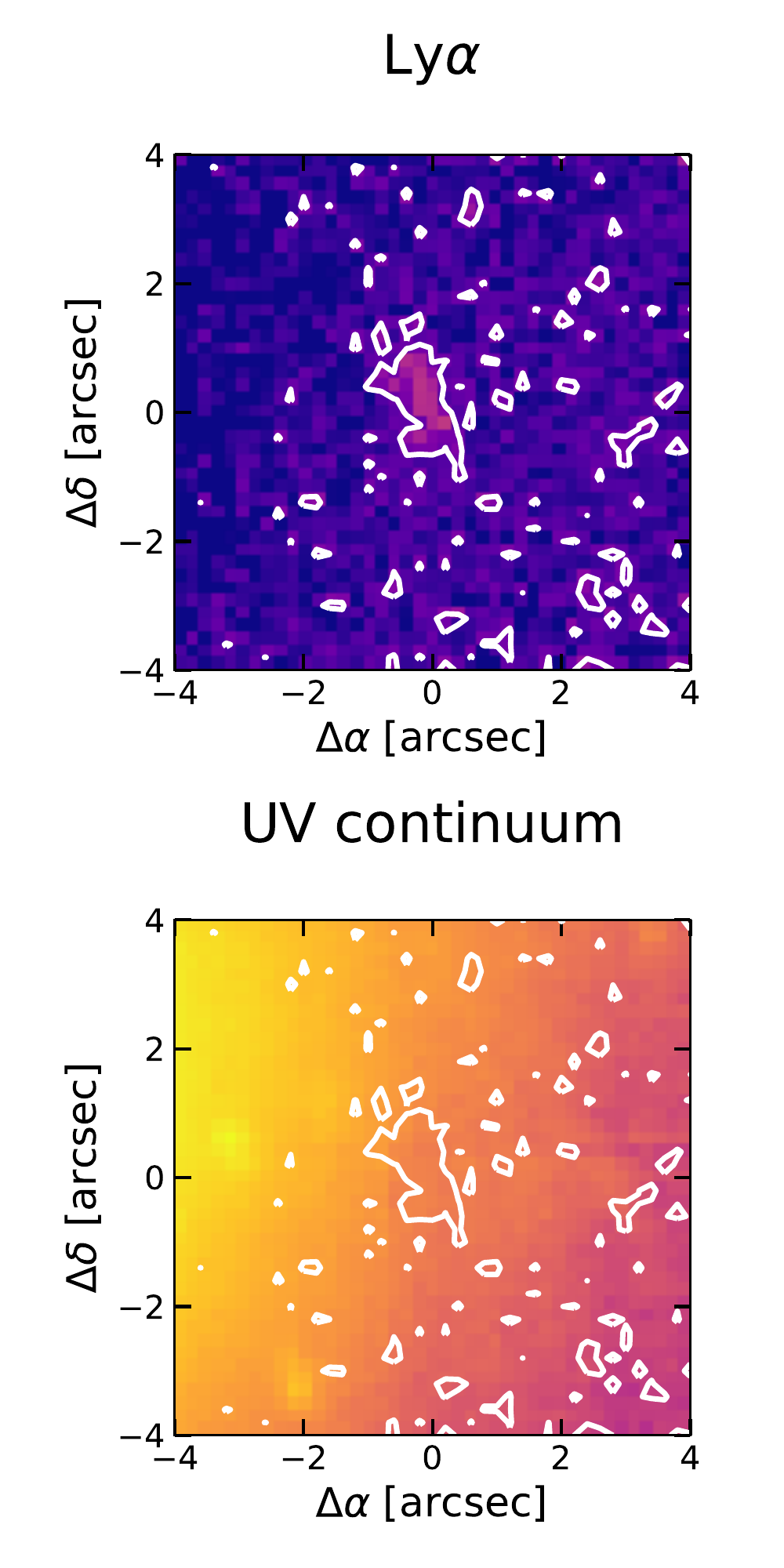}
\includegraphics[height=.32\textheight]{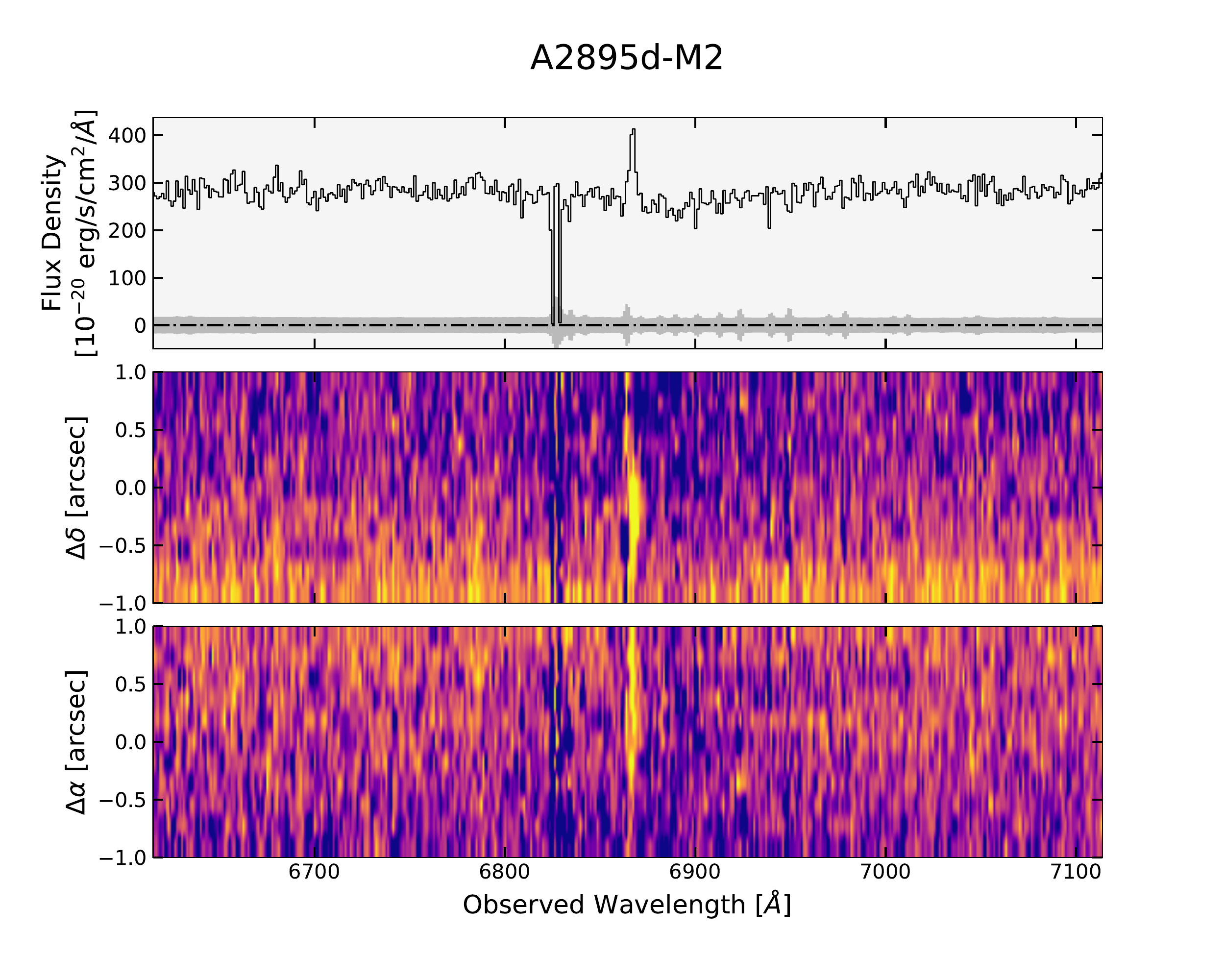}
\end{figure*}

\bigskip

\begin{figure*}
\includegraphics[height=.32\textheight]{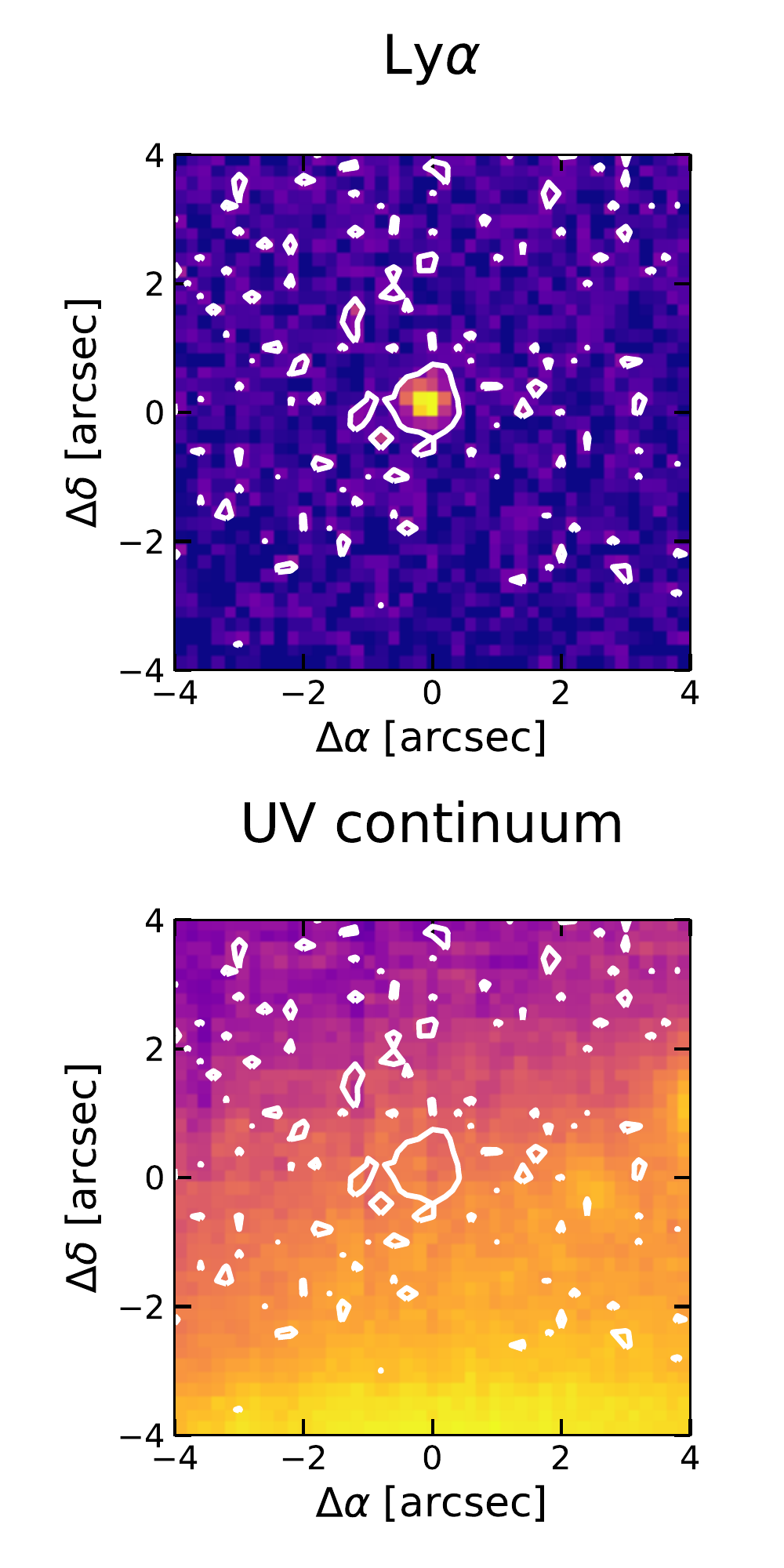}
\includegraphics[height=.32\textheight]{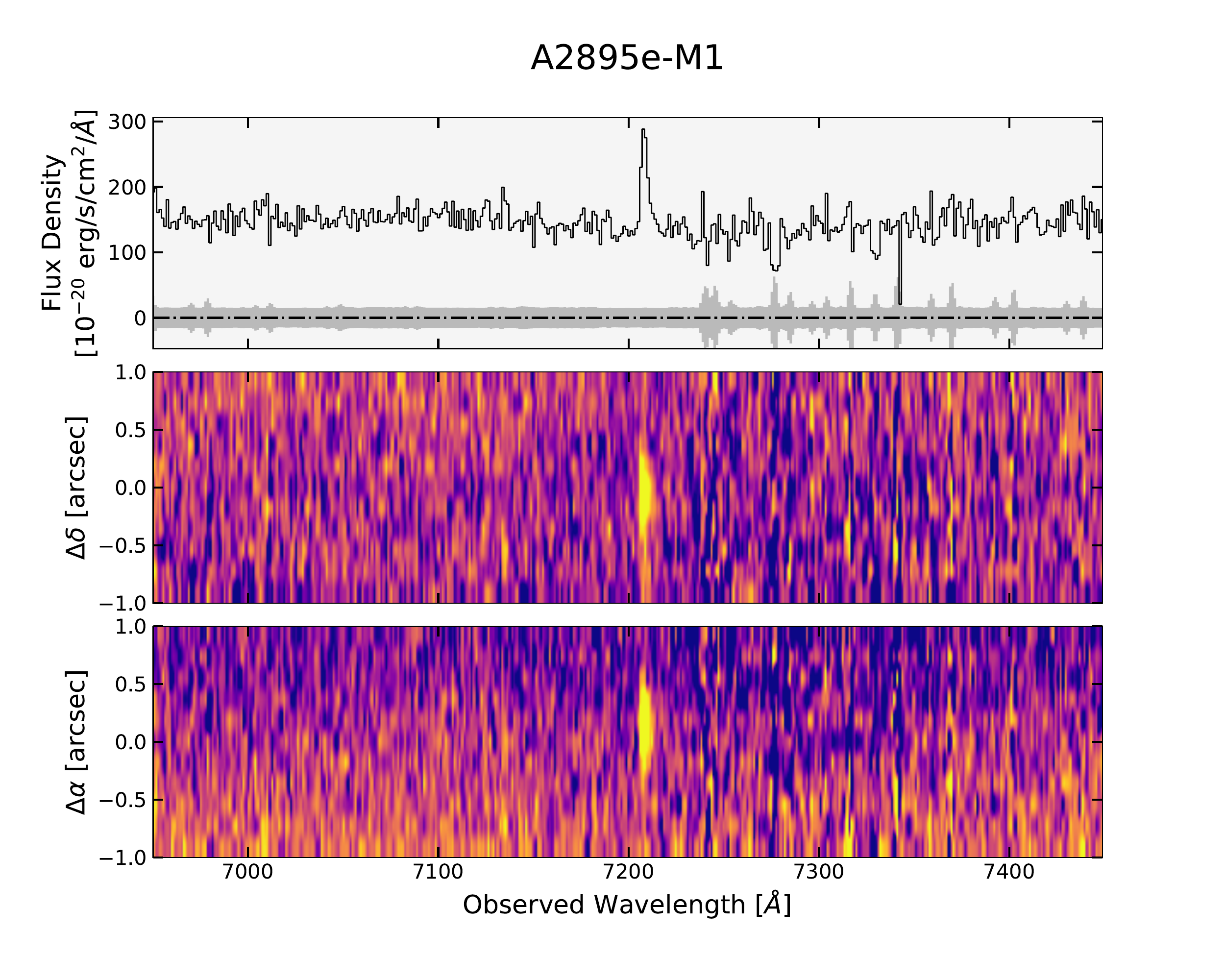}\\
\bigskip
\includegraphics[height=.32\textheight]{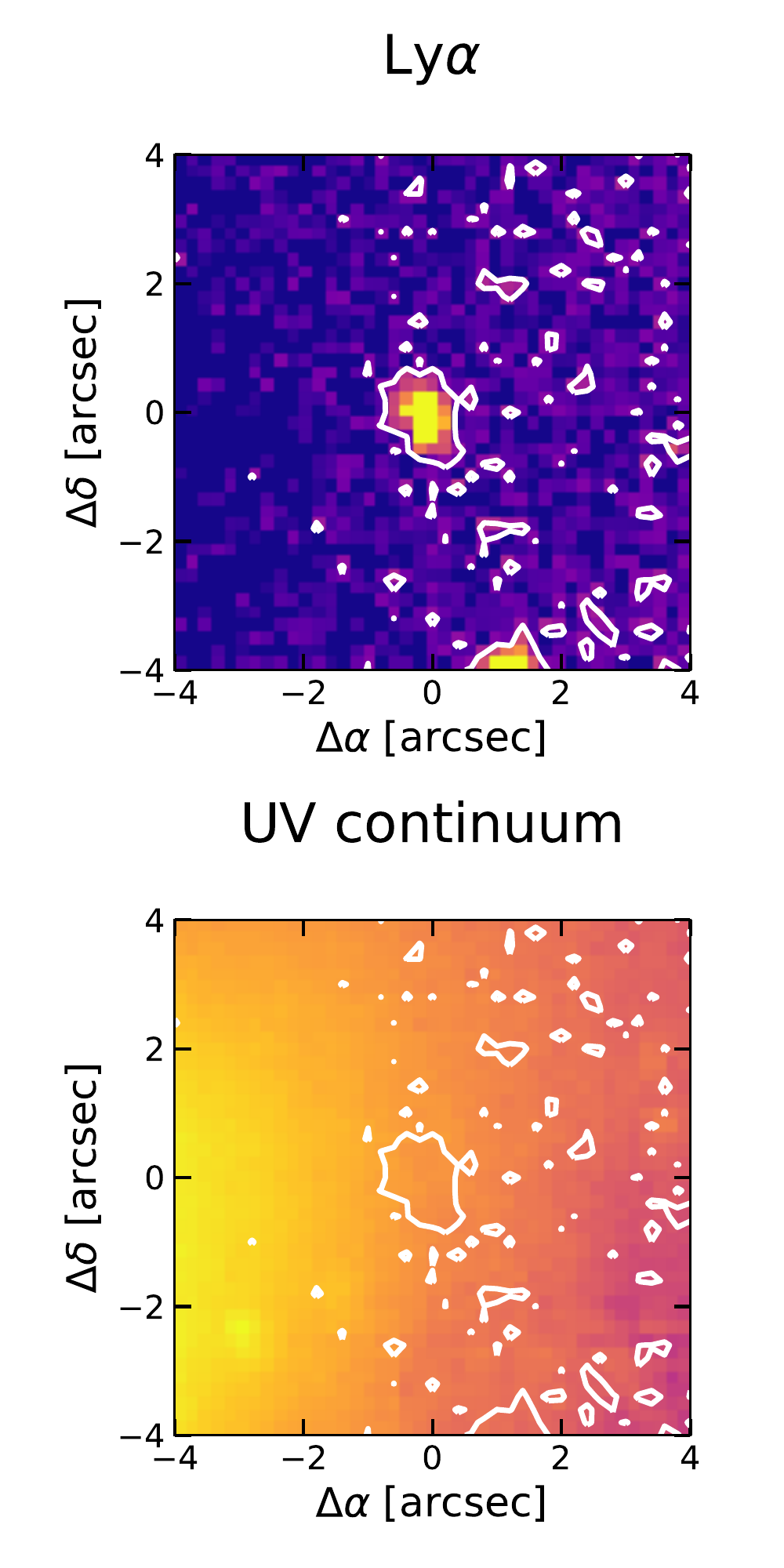}
\includegraphics[height=.32\textheight]{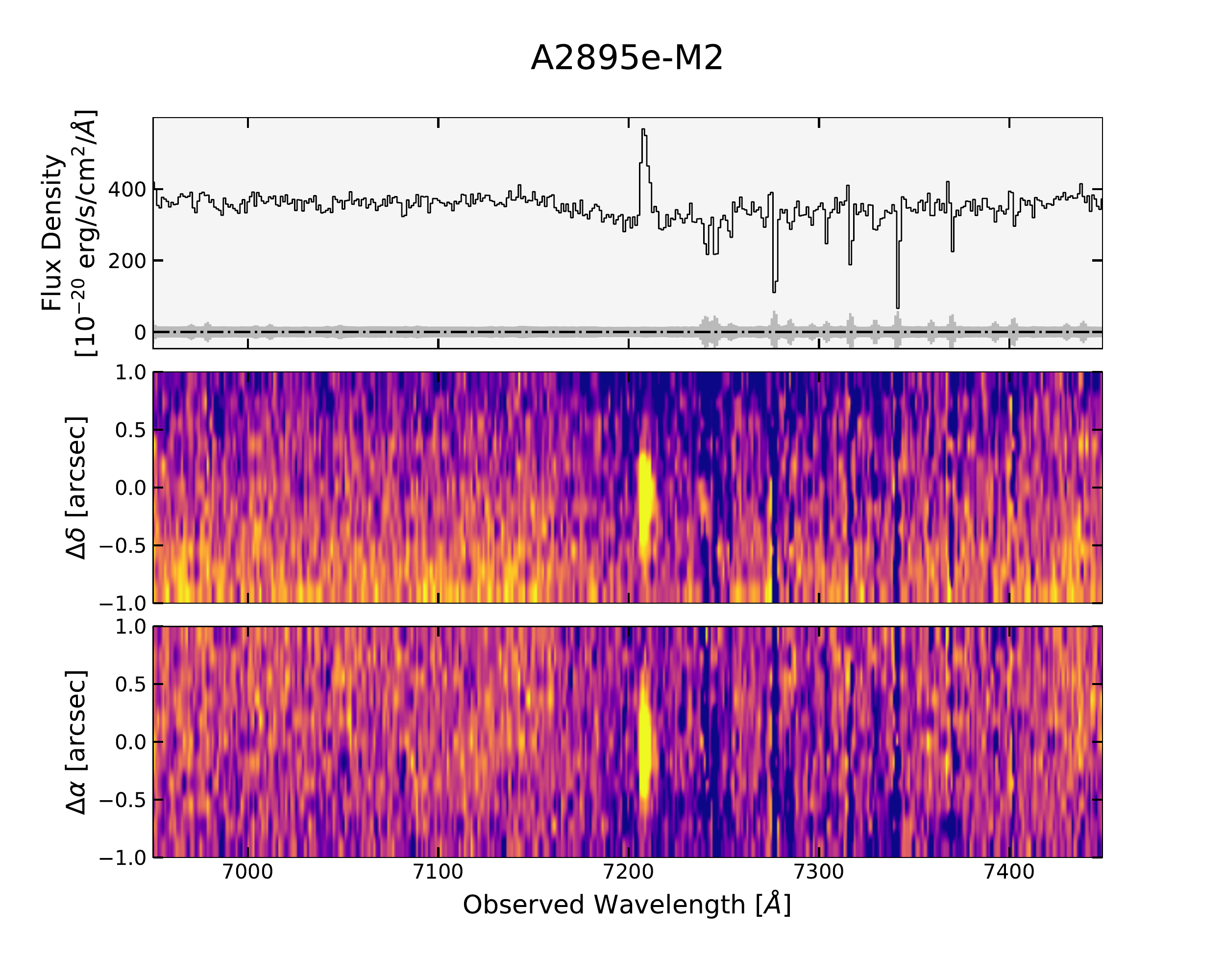}\\
\bigskip
\includegraphics[height=.32\textheight]{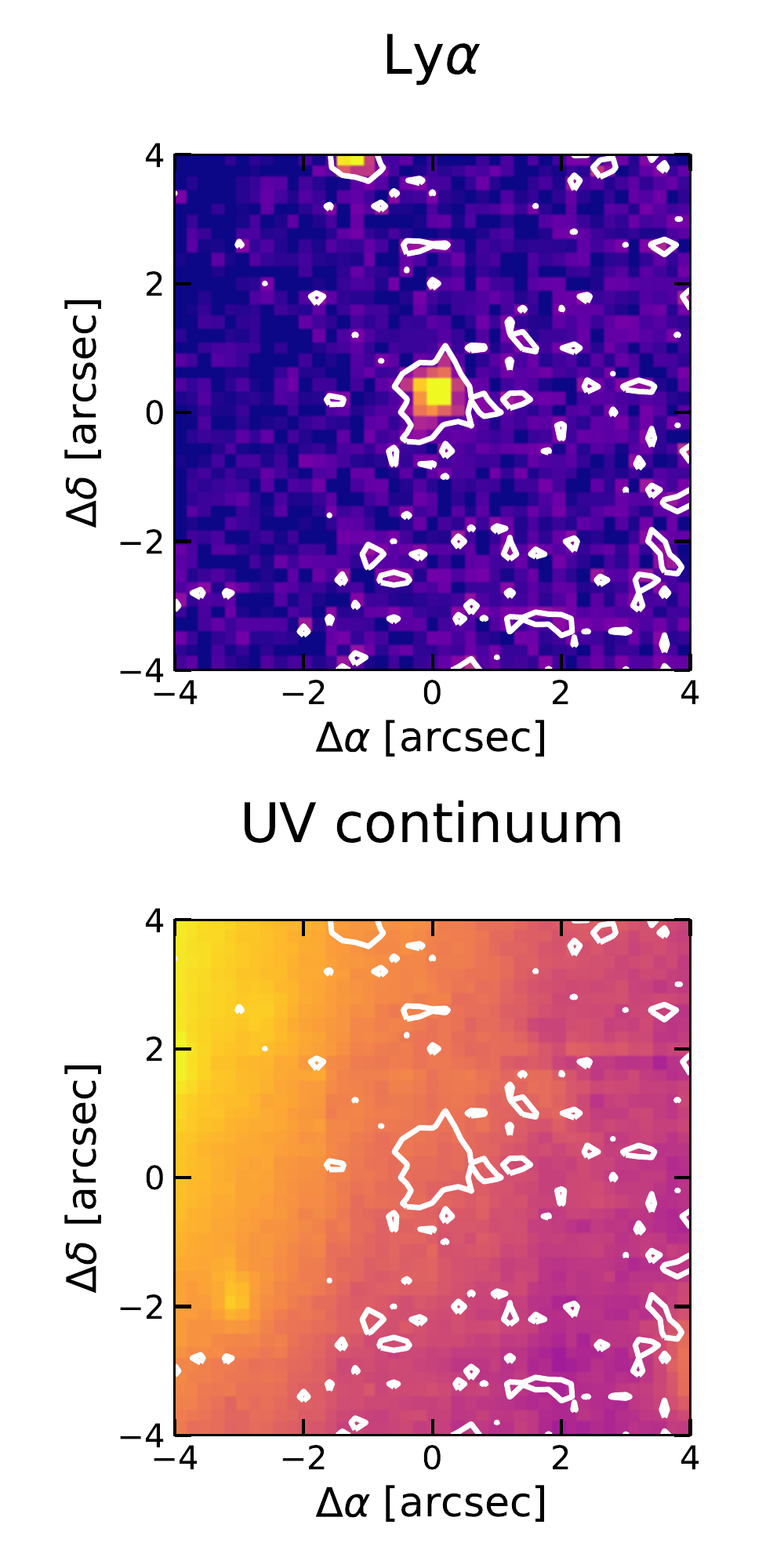}
\includegraphics[height=.32\textheight]{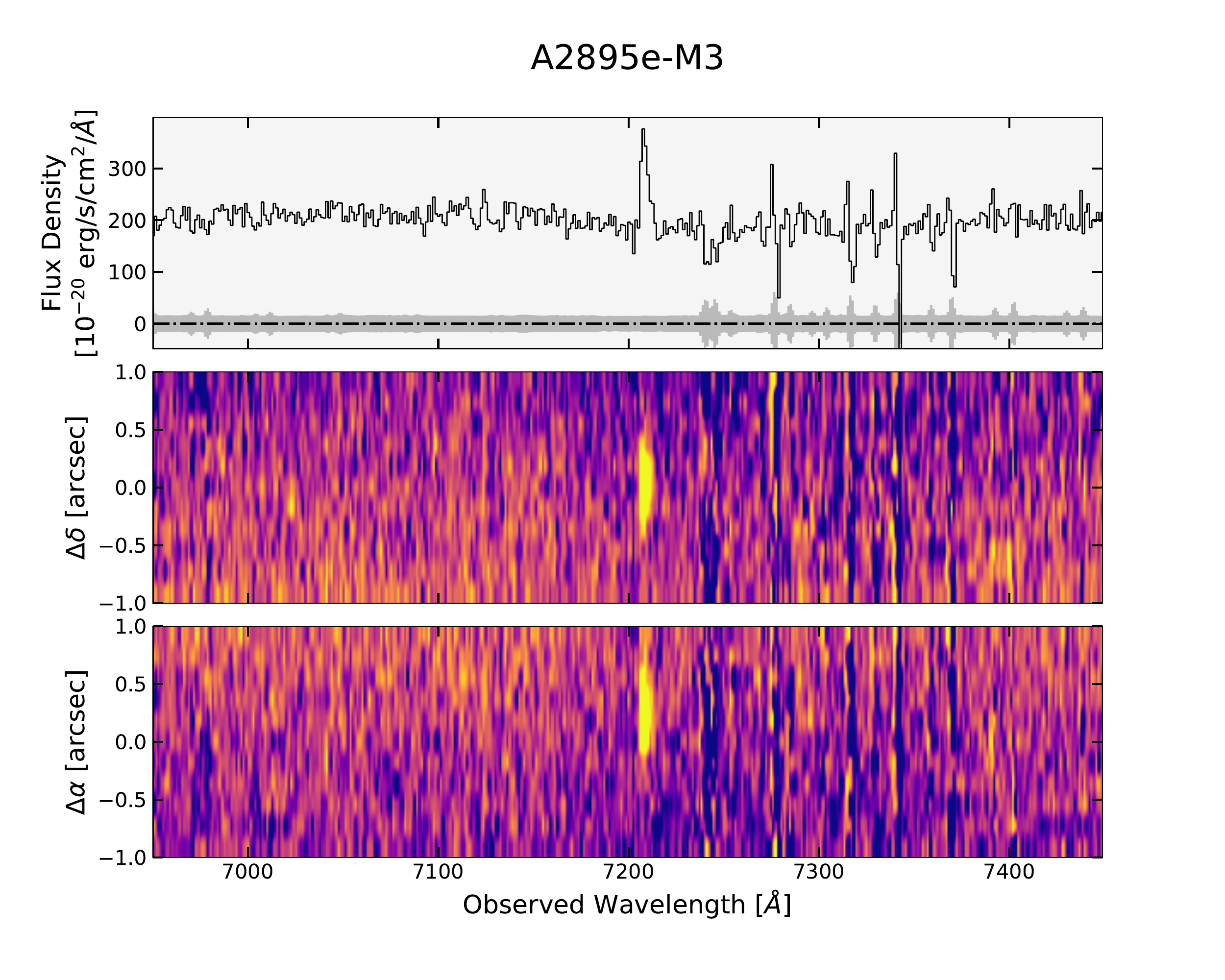}
\end{figure*}

\begin{figure*}
\includegraphics[height=.32\textheight]{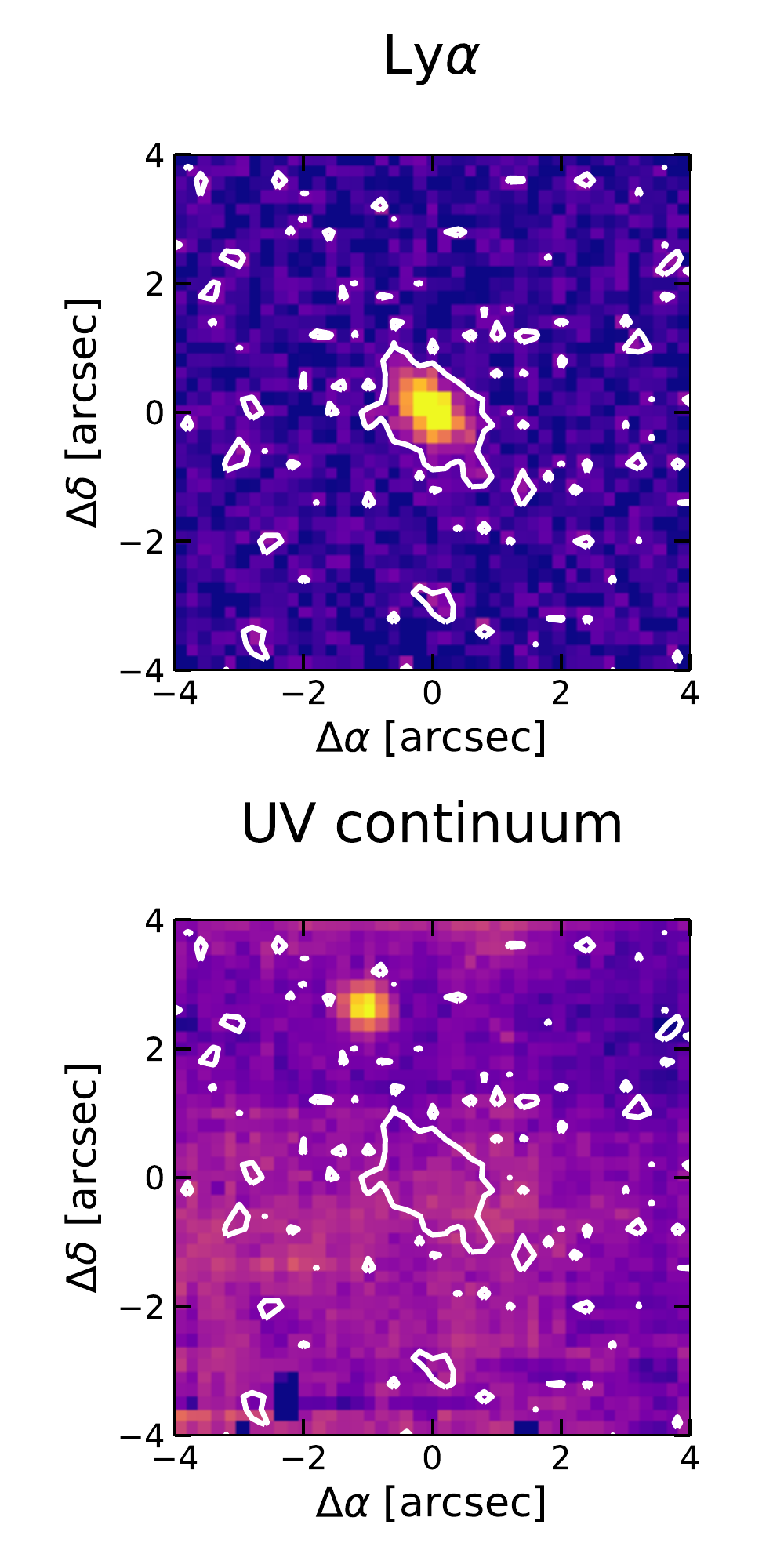}
\includegraphics[height=.32\textheight]{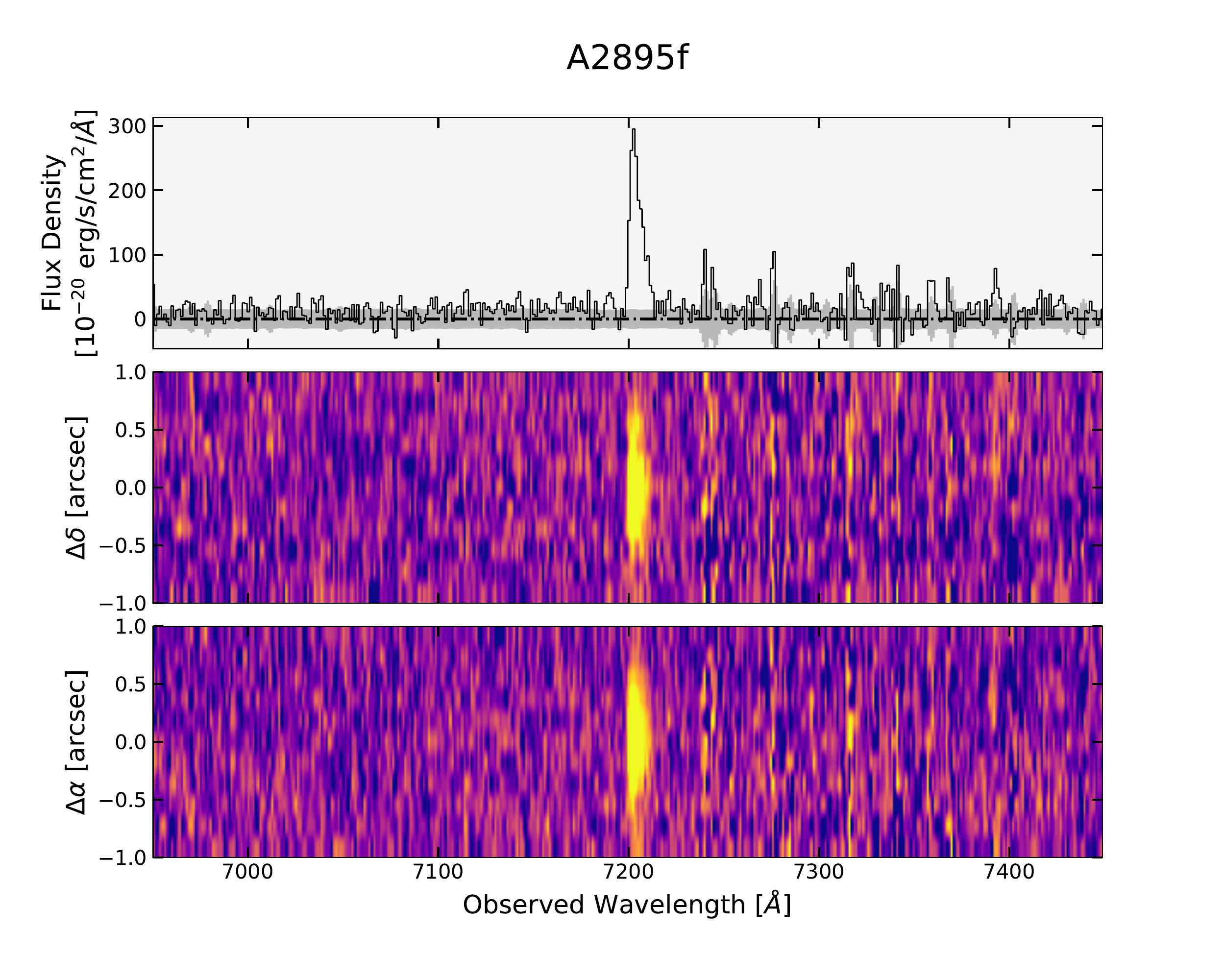}
\end{figure*}

\bigskip

\begin{figure*}
\includegraphics[height=.32\textheight]{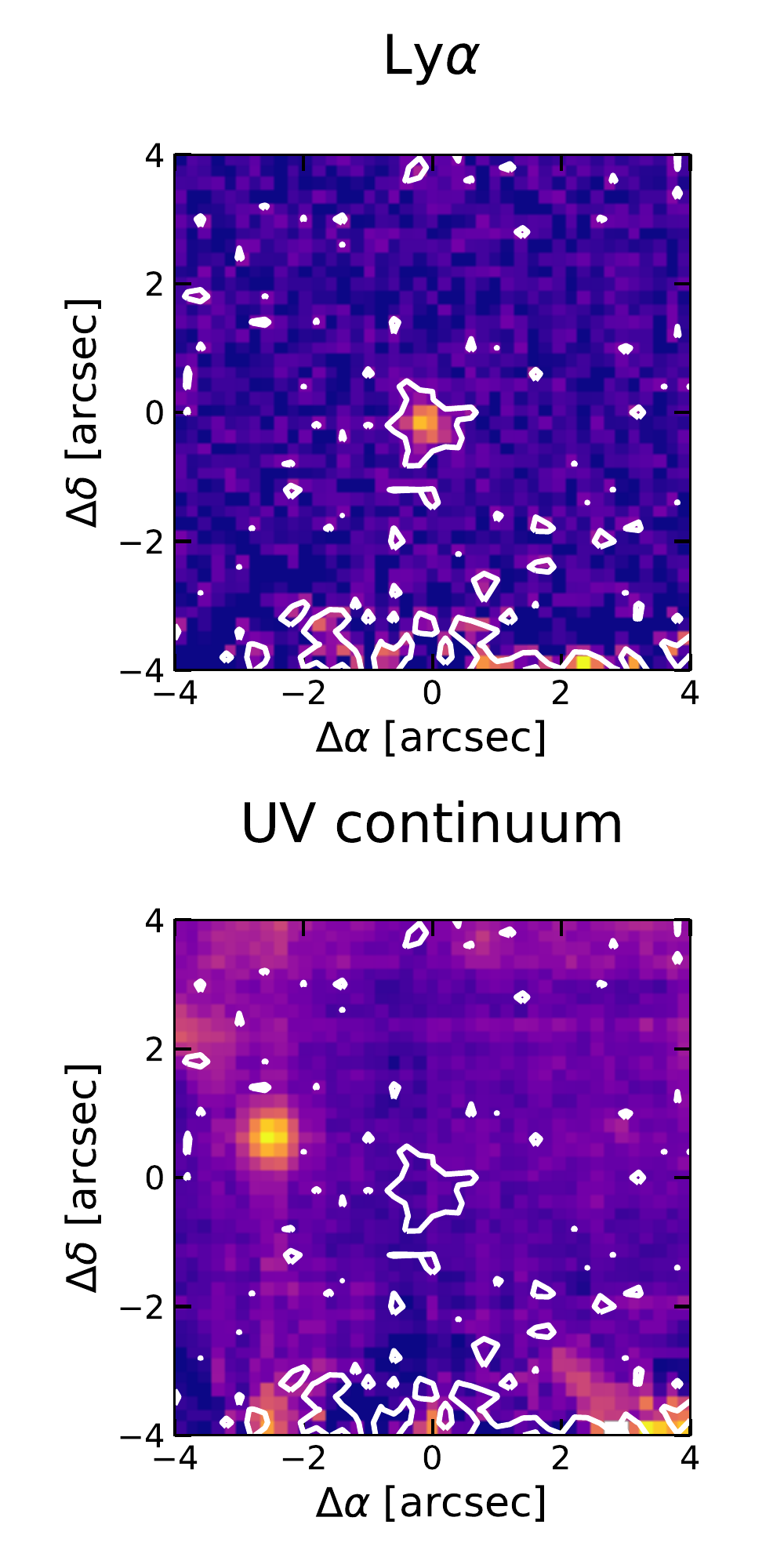}
\includegraphics[height=.32\textheight]{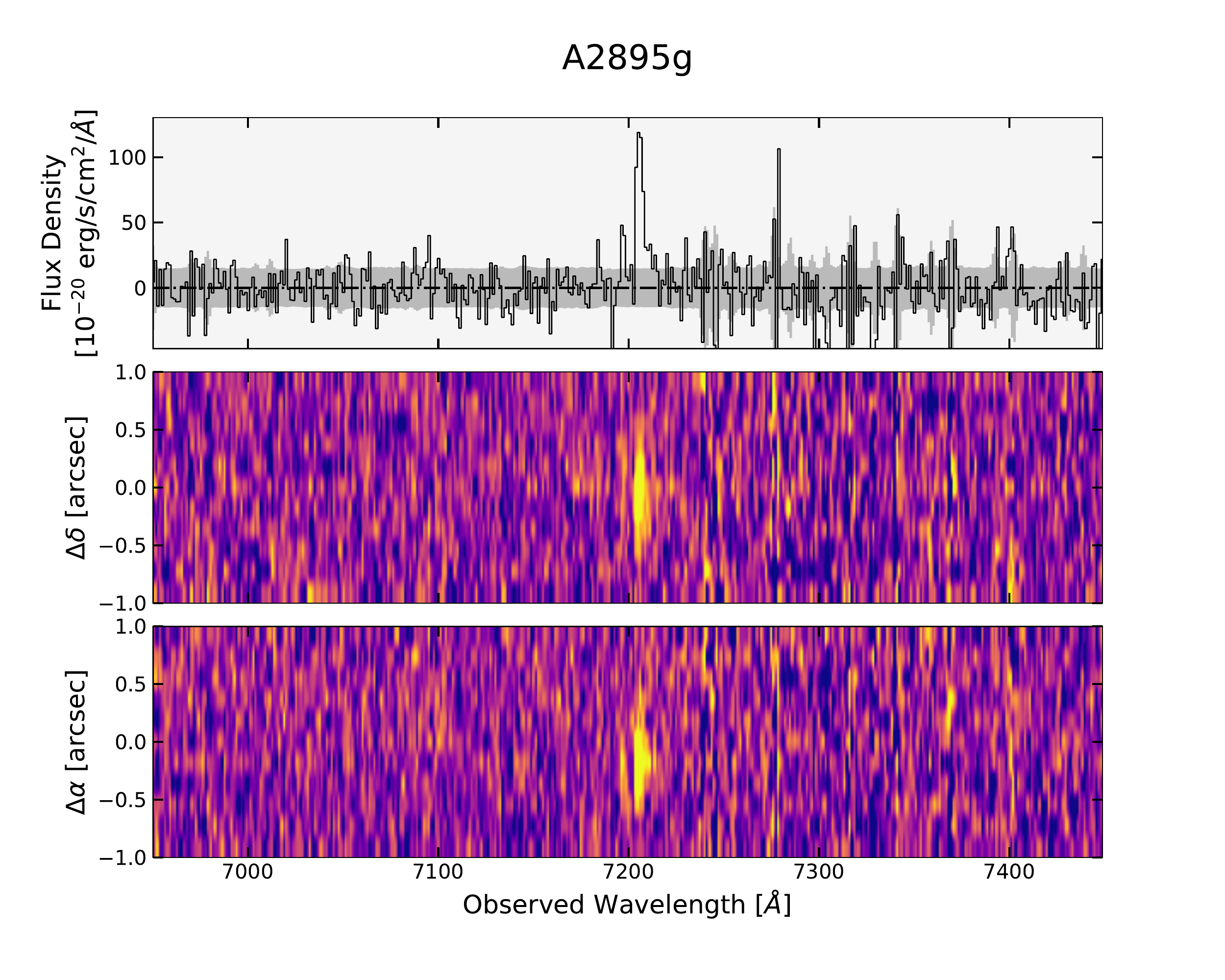}
\end{figure*}


\bsp	
\label{lastpage}
\end{document}